\documentclass[aps,rmp,twocolumn,nofootinbib]{revtex4}

\usepackage{amsmath}
\usepackage{epsfig}
\newcommand{\bea}{\begin{eqnarray}}
\newcommand{\eea}{\end{eqnarray}}
\newcommand{\be}{\begin{equation}}
\newcommand{\ee}{\end{equation}}

\begin{document}

\title{{\it Colloquium}: The spin-incoherent Luttinger liquid}

\author{Gregory A. Fiete}
\email{fiete@caltech.edu}
\affiliation{Department of Physics, California Institute of Technology, MC 114-36, Pasadena, California 91125, USA\\ and Kavli Institute for Theoretical Physics, University of California, Santa Barbara, California 93106, USA}

\begin{abstract}

In contrast to the well known Fermi liquid theory of three dimensions, interacting one-dimensional and quasi one-dimensional systems of fermions are described at low energy by an effective theory known as Luttinger liquid theory.  This theory is expressed in terms of  collective many-body excitations that show exotic behavior such as spin-charge separation.  Luttinger liquid theory is commonly applied on the premise that ``low energy" describes both the spin and charge sectors.  However, when the interactions in the system are very strong, as they typically are at low particle densities, the ratio of spin to charge energy may become exponentially small.  It is then possible at very low temperatures for the energy to be low compared to the characteristic charge energy, but still high compared to the characteristic spin energy.  This energy window of near ground-state charge degrees of freedom, but highly thermally excited spin degrees of freedom is called a spin-incoherent Luttinger liquid. The spin-incoherent Luttinger liquid exhibits a higher degree universality than the Luttinger liquid and its properties are qualitatively distinct.  In this colloquium  I detail some of the recent theoretical developments in the field and describe experimental indications of such a regime in gated semiconductor quantum wires.

\end{abstract}

\date{\today}
\maketitle
\tableofcontents

%73.21.-b Electron states and collective excitations in multilayers, quantum
%wells, mesoscopic, and nanoscale systems
%71.10.Pm Fermions in reduced dimensions
%73.21.Hb Quantum Wires
%73.23.Hk Coulomb blockade; single-electron tunneling

%71.10.Pm Fermions in reduced dimensions
%71.27.+a Strongly correlated electron systems;heavy fermions
%73.21.-b Electron states and collective excitations in multilayers, quantum
%wells, mesoscopic, and nanoscale systems
\section{Introduction}
\label{sec:intro}

This colloquium is about the properties of strongly interacting fermions in one spatial dimension in a certain window of energies, but before I come to the details of our main topic, it is worthwhile to recall what we ``know" about interacting fermions.  First, fermions obey Pauli exclusion statistics--no two particles can occupy the same state, i.e. have the same set of quantum numbers.  The simplest case to consider is a non-interacting system of fermions.  Often one is interested in the ground state and the nature of the lowest lying excitations above the ground state.  If we have have non-interacting fermions, the problem is a relatively simple one to solve.  We just find all the eigenstates of a single-particle Hamiltonian and then fill them (with one particle each because of the Pauli principle) starting from the lowest energy state until all $N$ electrons in the system occupy a state. By construction this is the ground state.  If $N$ is large, as it is for the number of electrons in a metal for example, one refers to the set of filled states as a ``Fermi sea" and the ``top'' of the sea is called a ``Fermi surface", which in one dimension actually consists of only two points in momentum space.  This sea structure is a direct consequence of the Pauli exclusion principle and our non-interacting particle assumption.  The lowest lying excitations are also easy to find: We just take a particle near the ``surface" and move it above the surface (since it cannot be moved below because all those states are filled by construction).  This process leaves a ``hole" in the Fermi sea and a particle excited above the Fermi sea.  Naturally, such excitations are called particle-hole excitations.

Now suppose we make the system more realistic by remembering that real fermions interact. The effect of these interactions remarkably turns out to depend on the spatial dimension being considered. In three dimensions the Fermi surface miraculously survives and the low-energy particle and hole excitations are very much like those in the non-interacting case only with a renormalization of their mass, and they acquire a lifetime inversely proportional to the square of the energy relative to the Fermi surface energy.  The theory describing this situation is called Fermi liquid theory and it has been successfully applied to liquid $^3$He and many metals \cite{Vollhardt:rmp84}.

On the other hand, in one spatial dimension, the Fermi surface does not survive (in the sense that no low energy excitations of single-particle character exist, although the Fermi wavevector $k_F$ remains a special value) and a new state is born--the Luttinger liquid. The Luttinger liquid contains only collective many-body excitations which can be separated into spin and charge sectors that propagate with different collective mode velocities. Note that this is remarkably different from the intuition we have from a single electron where the spin and charge are ``tied'' together. The details of how and why spin-charge separation occur are very well understood and there are a number of excellent sources on the topic.\footnote{See for example, \textcite{Voit:rpp95,Chang:rmp03,Giamarchi}.} We will not need those technical details here.

I believe the easiest way to understand the physics of the spin-incoherent Luttinger liquid is to see what goes ``wrong'' with a Luttinger liquid when the temperature is large compared to the characteristic spin energy, but still small compared to the characteristic charge energy.\footnote{This wide separation of spin and charge energy scales requires strong interactions and is discussed in Sec.~\ref{sec:Wigner}.}  For that we need to be familiar with a few of the central results of Luttinger liquid theory itself.  An excellent place to illustrate the consequences of spin-charge separation and get a flavor of the differences between the Luttinger liquid and the spin-incoherent Luttinger liquid is the single-particle Greens function.  We will come to what goes ``wrong'' with a Luttinger liquid when the temperature is higher than the spin energy later in this colloquium.

The single-particle Greens function in imaginary time $\tau=it$  is ${\cal G}_\sigma(x,\tau)=-\langle T_\tau\psi_\sigma(x,\tau)\psi_\sigma^\dagger(0,0)\rangle$ where $\psi_\sigma^\dagger/\psi_\sigma$ is the creation/annihilation operator for a fermion of spin projection $\sigma$ along the z-axis and $T_\tau$ is the $\tau$ ordering operator.  For a Luttinger liquid, one finds\footnote{The formula below assumes $SU(2)$ symmetry in the spin sector.}
\be
\label{eq:G_LL}
{\cal G}^{\rm LL}_\sigma(x,\tau)\sim \frac{1/\sqrt{v_s\tau-ix}}{(x^2+v_c^2\tau^2)^{\gamma_{K_c}}}\frac{e^{ik_Fx}}{\sqrt{v_c\tau-ix}}+c.c. ,
\ee
where $\gamma_{K_c}=(K_c+K_c^{-1}-2)/8\ge 0$, $v_s$ is the spin velocity, $v_c$ is the charge velocity, and the Fermi wavevector $k_F \equiv \pi/(2a)$ where $a$ is the interparticle spacing.  The single particle Greens function \eqref{eq:G_LL} clearly illustrates the effects of spin-charge separation when $v_s\neq v_c$: when a particle is added at (0,0) the charge and spin components propagate with different velocities.  Physically, this occurs because the Hamiltonian for the interacting 1-d system separates at low energies to $H=H_s+H_c$ where $H_s$ is the spin Hamiltonian and $H_c$ is the charge Hamiltonian. Note the branch cut structure of the singularities in \eqref{eq:G_LL}. The absence of a simple pole is related to the absence of quasi-particle excitations in a Luttinger liquid.  When the fermions are non-interacting one has $K_c=1$ and $v_s=v_c$, and the non-interacting Greens function is recovered from \eqref{eq:G_LL} along with the simple pole structure characteristic of quasi-particle excitations.

By contrast, in the spin incoherent case (defined as the regime where the spin energy $E_{\rm spin}\approx \hbar v_s/a \ll k_B T \ll \hbar v_c/a \approx E_{\rm charge}$ where $T$ is the temperature, $k_B$ is Boltzmann's constant, and $\hbar$ is Planck's constant divided by $2\pi$) we have
\be
\label{eq:G_SILL}
{\cal G}^{\rm SILL}_\sigma(x,\tau)\sim \frac{e^{-2k_F|x|(\ln 2/\pi)}}{(x^2+v_c^2\tau^2)^{\Delta_{K_c}}} \frac{e^{i(2k_F x -\varphi^+_{K_c})}}{v_c\tau-ix}+c.c.,
\ee
where $\Delta_{K_c}\neq\gamma_{K_c}$ and $\varphi_{K_c}$ are given in Sec.~\ref{sec:non_cons}.  In the spin-incoherent regime the Hamiltonian is still spin-charge separated and many features of the Greens function \eqref{eq:G_SILL} are reminiscent of  \eqref{eq:G_LL}.  However, there are important differences.  First note the exponential decay in \eqref{eq:G_SILL} which replaces the square root branch cut from the spin mode in the Luttinger liquid.  The exponential factor is distinct from the Luttinger liquid case \eqref{eq:G_LL} in which only power laws appear and can be said to put the spin-incoherent Luttinger liquid in a different ``universality class'' from the Luttinger liquid.  Second, note that the spin velocity $v_s$ has dropped out of \eqref{eq:G_SILL} implying that the spin degrees of freedom are non-propagating. It turns out that the absence of $v_s$ in the Greens function is a specific case of a more general ``super universal'' spin physics that occurs in the spin-incoherent regime in which no parameter of the spin Hamiltonian enters the correlation functions. As a result, the correlations are {\em completely independent of} $H_s$ and this implies a higher degree of universality in the spin-incoherent Luttinger liquid than in the Luttinger liquid since $H_s$ will not affect any observable properties of the system in this regime.  Third, note the shift $k_F \to 2k_F$ in the oscillating wavevector of the Green's function.  This shift comes from the spin-incoherence itself and can also be seen in the momentum distribution function $n(k)$ \cite{Cheianov:cm04}.

Another quantity which illustrates a remarkable difference between the Luttinger liquid and the spin-incoherent Luttinger liquid is the frequency dependence of the tunneling density of states (which can be derived from the imaginary part of the retarded Greens function).  For a Luttinger liquid one finds
\be
\label{eq:tdos_LL}
A^{\rm LL}(\omega)\sim \omega^{(K_c+K_c^{-1}-2)/4},
\ee
and for a spin-incoherent Luttinger liquid
\be
A^{\rm SILL}(\omega)\sim \omega^{(1/4K_c)-1}/\sqrt{|\log(\omega)|}.
\ee
As with the single-particle Greens function there is a qualitative difference in the tunneling density of states:  The Luttinger liquid always exhibits a suppression of the tunneling as $\omega \to 0$, while the spin-incoherent Luttinger liquid will show a divergence\footnote{The divergence will only be seen for $\hbar \omega \gtrsim k_BT$.  For smaller $\omega$ the divergence is cut off \cite{Matveev:prl06}.} for decreasing $\omega$ if $K_c>1/4$, which will be the case for many systems exhibiting the spin-incoherent Luttinger liquid phase.  In addition there are logarithmic corrections which turn out to appear quite generally in a whole class of quantities (see Sec.\ref{sec:non_cons}). There are many other distinct signatures of the spin-incoherent Luttinger liquid that appear in transport, interference experiments, tunneling, the Fermi-edge singularity, Coulomb drag, and noise that we will detail in the following sections of this colloquium.  

It is worth noting that while certain aspects of the spin-incoherent equal time Greens function were computed by \textcite{Berkovich:jpa91}, the full space-time dependence of the spin-incoherent Luttinger liquid regime appears to have been first studied in two beautiful papers--one by \textcite{Matveev:prl04} as a possible explanation for the 0.7 conductance feature in quantum point contacts and one by \textcite{cheianov03} in which they investigated the single-particle Greens function in the spin-incoherent regime of the infinite-$U$ Hubbard model.  All of these works have proved invaluable to establishing the field and to subsequent progress in it.

In the remainder of the colloquium I will develop some of the relevant theory to describe the spin-incoherent Luttinger liquid.  Since the spin-incoherent Luttinger liquid requires very strong interactions, a useful starting point is to model the fermions as a fluctuation Wigner solid.  This approach provides a convenient conceptual framework for understanding the most important aspects of the spin-incoherent Luttinger liquid.  For reasons that will become clear as we proceed, the features of the spin-incoherent Luttinger liquid fall into two categories depending on whether the relevant correlation functions are particle non-conserving (such as the single-particle Greens function) or particle conserving (such as the density-density correlation function). I will cover these issues in detail and illustrate the main points with examples in each case.  The colloquium will close with a brief discussion of experimental indications of the spin-incoherent Luttinger liquid in semiconductor quantum wires and some of the outstanding problems in the field.

\section{Fluctuating Wigner Solid model}
\label{sec:Wigner}

\begin{figure}[h]
\includegraphics[width=.95\linewidth,clip=]{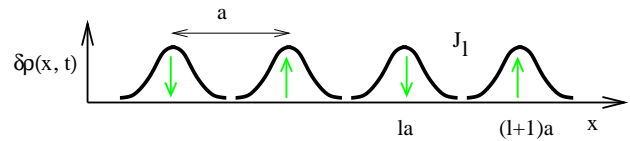}
\caption{\label{fig:solid_schematic_rev} (color online) Schematic of a fluctuating Wigner solid model. Shown is a snap shot in time of the density modulations $\delta\rho(x,t)$ of a fluctuating Wigner solid  in one dimension of mean particle spacing $a$. A cartoon of the antiferromagnetic spin sector is shown with the green arrows representing spin orientation on the Wigner solid ``sites'' given by $la$.  There is an antiferromagnetic exchange $J_l$ between the $l^{th}$ and $(l+1)^{th}$ site.  The lattice has a stiffness characterized by the frequency $\omega_0 \gg J_l/\hbar$. }
\end{figure}

The spin-incoherent Luttinger liquid requires $E_{\rm spin} \ll E_{\rm charge}$ and this separation of spin and charge energy scales only occurs for strong interactions. The discussion in this section attempts to explain clearly why strong interactions are necessary and how to simply model this situation. Consider a one-dimensional gas of spin-1/2 fermions.  The typical kinetic energy  $K.E.\sim \frac{\hbar^2 k_F^2}{2m^*}$ where $m^*$ is the effective mass, and the typical potential energy  $P.E.\sim \frac{e^2}{\epsilon a}$ where $\epsilon$ is the dielectric constant and $e$ is the charge of the fermion. Since $k_F \propto n =1/a$, we see immediately that $\frac{P.E.}{K.E.}\sim \frac{1}{na_B}\sim r_s$, where $a_B\equiv \frac{\epsilon\hbar^2}{m^*e^2}$ is the Bohr radius of the material and $r_s\equiv1/(2na_B)$ is a dimensionless parameter describing how strong the potential energy is relative to the kinetic energy.  Obviously, the larger $r_s$ is, the more solid-like the electron gas. (See Fig.~\ref{fig:solid_schematic_rev} for a schematic.)  The solid phase is actually believed to be obtained in two and three dimensions for sufficiently large $r_s$ \cite{Tanatar:prb89}.  However, in one dimension the quantum fluctuations are strong enough to destroy the long range order, even for long range Coulomb interactions \cite{Schulz:prl93}.  Nevertheless, a ``fluctuating'' Wigner solid still provides a useful starting point for strongly interacting fermions in one dimension and I will use such a model throughout the remainder of this colloquium.

\subsection{Effective Hamiltonian}
A fluctuating Wigner solid is mathematically equivalent to a harmonic chain\footnote{An alternative description of a Wigner solid has been discussed by \textcite{Novikov:prl05,Novikov:prb05}.} with Hamiltonian
\begin{equation}
\label{eq:H_chain}
H_{\rm chain}=\sum_{l=1}^N  \frac{p_l^2}{2m^*}+\frac{m^*\omega_0^2}{2}(u_{l+1}-u_l)^2\;,
\end{equation}
where $p_l$ is the momentum of the $l^{th}$ fermion, $u_l$ the displacement from equilibrium of the $l^{th}$ fermion, and $\omega_0$ the frequency of local fermion displacements.\footnote{When the physical system of interest is electrons in a quantum wire, $\omega_0$ will depend on the density, the width of the quantum wires, the dielectric constant of the material, and the distance to a nearby metallic gate \cite{Glazman:prb92,Hausler:prb02}.}  The position of the fermions along the chain are given by
\begin{equation}
x_l=l a +u_l\;,
\end{equation}
where $a$ is the mean spacing of the fermions as before.  Since we are dealing with a quantum system, we can canonically quantize the harmonic chain by imposing the commutation relation $[u_l,p_{l'}]=i\hbar\delta_{ll'}$.

Fermions also have a spin degree of freedom, and in one dimension very general considerations \cite{lieb62} result in antiferromagnetic interactions.  The simplest situation is when the exchange is of the nearest-neighbor type in the Wigner solid.  In this case, the Hamiltonian of the spin sector takes the form
\begin{equation}
\label{eq:H_s}
H_s=\sum_l J_l {\vec S_l}\cdot{\vec S_{l+1}}\;,
\end{equation}
where $J_l$ is the nearest neighbor exchange between the $l^{th}$ and $(l+1)^{th}$ sites in the lattice. (See Fig.~\ref{fig:solid_schematic_rev}.) The spin Hamiltonian \eqref{eq:H_s} is the generic low-density form for $SU(2)$ symmetric interactions regardless of whether the interactions between electrons are zero range, as in the Hubbard model, or long range unscreened Coulomb interactions \cite{Ogata:prb90,Matveev:prb04}.  Moreover, in the strongly interacting regime assumed here, one can argue on the general grounds that the exchange energy $J_l$ is exponentially suppressed relative to $\hbar \omega_0$, the characteristic charge energy.  Various approximations for $J_l$ have been elegantly discussed in a series of papers.\footnote{See for example, \textcite{Hausler:zpb96,Fogler_exch:prb05,Klironomos:prb05,Matveev:prb04}.}  The main physical point is that in one-dimension when the interactions are strong, two particles must tunnel through each other in order to ``exchange''.  It is just these processes that set the scale for $J_l$ which is exponentially small relative to $\hbar \omega_0$ because of the tunneling processes involved.\footnote{Although the physics we discuss here assumes low density, strongly correlated physics can also occur in very thin quantum wires at higher density \cite{Fogler:prb05,Fogler:prl05,Kindermann:prb07}.}

In this colloquium we will only be interested in energy scales small compared to the characteristic charge energy $\hbar \omega_0$, so we would like to find a simpler, low energy form for \eqref{eq:H_chain}.  To do so, we take a continuum limit and convert the sum over $l$ to an integral over $x$ \cite{Schulz:prl93,Matveev:prb04,Giamarchi}.  In order to make the low energy Hamiltonian look like the Hamiltonian familiar from Luttinger liquid theory, we also rescale the variables $u_l\to u(x)\equiv a \frac{\sqrt{2}}{\pi}\theta_c(x)$ and $p_l\to p(x) \equiv \frac{\hbar}{a\sqrt{2}}\partial_x\phi_c(x)$ which then satisfy the commutation relations $[u(x),p(x')]=\hbar [\theta_c(x),\partial_{x'}\phi_c(x')]=i \hbar \pi \delta(x-x')$ which is just the continuum version of $[u_l,p_{l'}]=i \hbar \delta_{ll'}$.  The resulting low energy charge Hamiltonian is
\be
\label{eq:H_c}
H_c=\hbar v_c\int \frac{dx}{2\pi}\left[\frac{1}{K_c}(\partial_x \theta_c(x))^2+K_c (\partial_x \phi_c(x))^2\right],
\ee
where $v_c=\omega_0a$ and $K_c=\frac{\pi\hbar}{2 am v_c}$.  For the rest of this colloquium, we will exclusively use \eqref{eq:H_c} to describe the charge sector of the spin-incoherent Luttinger liquid as we will only be interested in energies low compared to the characteristic charge energy, $E_{\rm charge}\approx \hbar \omega_0$.  The full Hamiltonian for all energies of interest is then
\be
\label{eq:H}
H= H_c +H_s,
\ee
with $H_c$ given by \eqref{eq:H_c} and $H_s$ given by \eqref{eq:H_s}.

\subsection{``Super universal'' spin physics}

Having obtained the relevant Hamiltonian we are now in a position to compute expectation values,
\be
\langle A \rangle = \frac{1}{Z}{\rm Tr}[e^{-\beta H}A],
\ee
where $A$ is any operator, $Z\equiv {\rm Tr}[e^{-\beta H}]$ is the partition function, and $\beta=(k_BT)^{-1}$. Since the Hamiltonian separates into separate spin and charge pieces, the trace can be evaluated for each independently. Consider the trace over the spin degrees of freedom.\footnote{I am assuming here for simplicity that $J_l=J$ independent of $l$, i.e., there is no magneto-elastic coupling.  In Sec.~\ref{sec:cons} we will see there are important effects resulting from the $l$ dependence of $J_l$.} In the spin-incoherent regime we have $E_{\rm spin} \ll k_B T$ which implies $e^{-\beta H_s}\approx 1$.  Therefore, {\em the spin Hamiltonian completely drops out of the trace}.  This is precisely the reason why no parameter of the spin Hamiltonian appeared in the single-particle Greens function \eqref{eq:G_SILL}, and one can see here that it is a very general feature of the spin-incoherent regime that expectation values (and correlation functions derived from them) are completely independent of $H_s$.  In this sense the spin physics exhibited in the regime  $E_{\rm spin} \ll k_B T$ is ``super universal'' and one can see, perhaps trivially, that the spin-incoherent Luttinger liquid has a higher degree of universality than a Luttinger liquid whose correlation functions will depend on $H_s$.  Note that even if $H_s$ is gapped the correlations will remain unaffected provided $E_{\rm spin} \ll k_BT$.

\subsection{Effective low-energy spin Hamiltonian}

Before moving on to the next section, we note the low-energy (for $k_BT\ll E_{\rm spin}$) form of \eqref{eq:H_s} is given by 
\be
\label{eq:H_s_low}
H_s^{\rm low}=\hbar v_s\int \frac{dx}{2\pi}\left[\frac{1}{K_s}(\partial_x \theta_s(x))^2+K_s (\partial_x \phi_s(x))^2\right],
\ee
where $v_s=Ja/\hbar$ is the spin velocity, and $K_s=1$ for an $SU(2)$ symmetric spin sector.  The bosonic fields satisfy $[\theta_\alpha(x),\phi_\beta(y)]=-i\frac{1}{2}\pi \delta_{\alpha,\beta}{\rm sgn}(x-y)$ where $\alpha,\beta=s$ or $c$.  In later sections we will make use of the low-energy form \eqref{eq:H_s_low} when we discuss how the Luttinger liquid breaks down when $k_BT$ approaches $E_{\rm spin}$ from below.  Any Luttinger liquid (with $v_s/v_c \ll 1$) at $T=0$ becomes a spin-incoherent Luttinger liquid  when $E_{\rm spin} \ll k_B T \ll E_{\rm charge}$.

\section{Particle non-conserving operators}
\label{sec:non_cons}§

Armed with the Hamiltonian \eqref{eq:H} one can calculate the properties of the spin-incoherent Luttinger liquid.  In this section, we will focus on results obtained from correlation functions of particle non-conserving operators, such as the particle creation or annihilation operator that appears in the single-particle Greens function.  These correlation functions will be evaluated in a first quantized path integral representation which makes clear why there is a distinction between particle conserving and non-conserving operators in the spin-incoherent regime. 

\subsection{Single-particle Greens function}
 \label{sec:Greens_gen}

\begin{figure}[h]
\includegraphics[width=.65\linewidth,clip=]{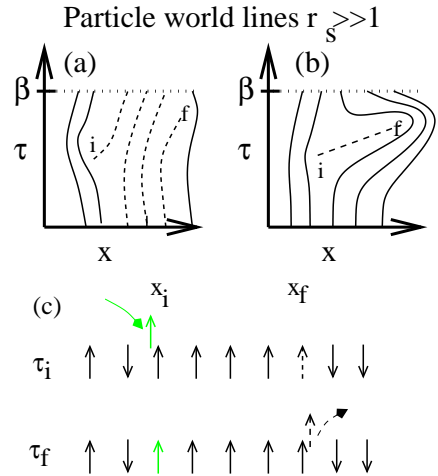}
\caption{\label{fig:worldlines} (color online) World lines for strongly interacting one dimensional spin-$S$ fermions for $E_{\rm spin} \ll k_B T \ll E_{\rm charge}$ in which the ``non-crossing'' approximation is made.  Particle trajectories in space and imaginary time are shown as curved lines.  The dashed lines represent the world line paths for creating a particle and removing it for large $x=x_f-x_i, \tau=\tau_f-\tau_i$.  The solid lines represent trajectories of other particles.  Because of the large action cost associated with the trajectories in (b), at low energies (relative to $E_{\rm charge}$) a process like that shown in (a) where world lines wrap around from $\tau=\beta$ to $\tau=0$ will dominate.  Such a process, however, requires that all dashed world lines have the same spin.  For $E_{\rm spin} \ll k_B T$ in zero magnetic field, this occurs with probability $(2S+1)^{-|N|}$ as discussed in the text.  Physically, the process shown in (a) is equivalent to adding an electron at $(x_i,\tau_i)$ then ``pushing'' all the electrons to the right and removing the electron at $(x_f,\tau_f)$ as shown in (c).  In order for the final spin configuration to ``look'' the same as the original one, all spins must be aligned between the initial and final point.}
\end{figure}

 The single-particle Greens function in imaginary time $\tau=it$  is 
\be
\label{eq:G_def}
{\cal G}_\sigma(x,\tau)=-\frac{1}{Z}{\rm Tr}[e^{-\beta H}\psi_\sigma(x,\tau)\psi_\sigma^\dagger(0,0)], \hspace{.4cm}\tau>0,
\ee
 where $\psi_\sigma^\dagger/\psi_\sigma$ is the creation/annihilation operator for a fermion of spin projection $\sigma$ along the z-axis and $Z\equiv {\rm Tr}[e^{-\beta H}]$ is the partition function as before. We evaluate the trace using a {\em first quantized path integral representation} of ${\cal G}_\sigma(x,\tau)$ \cite{Fiete:prl04}.  In this representation the trace is an integral over world lines (paths) of particle trajectories.\footnote{There are remarkable similarities between the analysis here and that for the gapped phases of 1d spin chains \cite{Damle:prl05,Rapp:prb06,Sachdev:prl97,Damle:prb97}.} Each configuration is weighted by a Euclidean action describing the ``deformation'' of world lines (see Fig.~\ref{fig:worldlines}) and a statistical factor reflecting the sign of the permutation of the fermions.

In evaluating the trace over the spin and charge degrees of freedom it is convenient to make maximum use of the hierarchy of energies relevant to the spin-incoherent Luttinger liquid: $E_{\rm spin}\ll k_B T \ll E_{\rm charge}$. A crucial part of our method of evaluating the trace is to make a {\em non-crossing approximation} in the space of world lines.  As explained earlier, $E_{\rm spin}$ is small because particle exchanges are rare due to the fact that particles must tunnel through each other when the interactions are strong.  Given $E_{\rm spin}$, this argument can then be turned around: the typical exchange (``crossing'') time $t_{\rm exch}\sim \hbar/E_{\rm spin}$ is very long. In particular, the exhange time is much longer than the thermal coherence time $t_{\rm coh}\sim \hbar/k_BT=\hbar \beta$ which is the amount of ``time'' a particle has to propagate in the imaginary time path integral formulation.  Therefore, on the time scale of the coherence time, no particles will exchange their positions.  We call this the ``non-crossing approximation'' and it restricts the possible world line trajectories to the types shown in Fig.~\ref{fig:worldlines}. It is worth emphasizing that this argument is completely general and exploits no feature of $H_s$ other than its characteristic energy $E_{\rm spin}$.

Having restricted the class of world lines that contribute to the Greens function in the spin-incoherent Luttinger liquid, a further subset can be selected that provide the dominant contribution to the Greens function.  These are the configurations shown in Fig.~\ref{fig:worldlines}(a) in which the trajectories of the added particle effectively ``wraps around'' the imaginary time  torus. These contributions minimize the action of the charge sector while obeying the non-crossing approximation. Contributions such as those shown in Fig.~\ref{fig:worldlines}(b) are exponentially suppressed in weight relative to those shown in Fig.~\ref{fig:worldlines}(a). In order for the trajectories to wrap around the imaginary time torus, the spin configuration of each world line between the initial and final point must be the same.  For a spin-$S$ fermion, this will occur with probability $(2S+1)^{-|N|}$ where $N$ is the number of world lines between the initial and final position.  Additionally, there is a permutation factor $(-1)^N$ coming from the fermi statistics.  Selecting this class of trajectories thus contrains the spins and is therefore equivalent to computing the trace over the spin degrees of freedom. The resulting dynamics therefore becomes effectively spinless and we are left with 
\bea
\label{eq:G_gen}
{\cal G}_\sigma(x,\tau)\sim \sum_{m=-\infty}^\infty \langle \delta(m-N(x,\tau))(-1)^m \hspace{1.5cm}\nonumber \\ \times (2S+1)^{-|m|}e^{i[\phi(x,\tau)-\phi(0,0)]}\rangle,
\eea
where the number of world lines $N(x,\tau)$ is allowed to fluctuate dynamically and the factor $e^{i\phi(x,\tau)}/e^{-i\phi(x,\tau)}$ annihilates/creates a particle at $(x,\tau)$. One must sum over all possible number of particles $m$ between the initial and final point.  

The remaining expectation value is computed over the charge degrees of freedom alone.  It is at this stage that we make use of the second part of the spin-incoherent Luttinger liquid energy hierarchy, $k_B T \ll E_{\rm charge}$, to compute the remaining average for $T\to 0$  using \eqref{eq:H_c}.  In order to compute the average over the charge degrees of freedom the number of particles $N(x,\tau)$ must be related to the fluctuating charge variables of the Hamiltonian \eqref{eq:H_c}. This relation is obtained simply by integrating up the fluctuating density between the intial and final space-time points \cite{Fiete:prl04}
\bea
\label{eq:N_with_theta}
N(x,\tau)&=&nx + \frac{\sqrt{2}}{\pi}(\theta_c(x,\tau)-\theta_c(0,0)) \nonumber \\
&=&nx + \frac{1}{\pi}(\theta(x,\tau)-\theta(0,0)),
\eea
where the spinless fields satisfy the same commutation relations as the original charge fields: $[\theta(x),\partial_{x'}\phi(x')]=[\theta_c(x),\partial_{x'}\phi_c(x')]=i \pi \delta(x-x')$.  We now turn to an evaluation of \eqref{eq:G_gen} for the cases of large and small $x$.

\subsubsection{${\cal G}_\sigma(x,\tau)$ for large $x$}

When $x$ is large, $N(x,\tau)$ is also large and little error is made in converting the sum in \eqref{eq:G_gen} to an integral, $\sum_m \to \int dm$. When this is done, the integral over $m$ is readily evaluated with the delta function and one obtains
\be
 \label{eq:G_gen_large_x}
{\cal G}_\sigma(x,\tau)\sim \langle(-1)^{N(x,\tau)} (2S+1)^{-|N(x,\tau)|}e^{i[\phi(x,\tau)-\phi(0,0)]}\rangle,
\ee
where the expectation value over the charge degrees of freedom is taken in the limit $T\to 0$.\footnote{All of the spin incoherent results presented in this section are computed at zero temperature in the charge sector. This is equivalent to taking the order of limits $J\to 0$, then $T \to 0$.} In order to deal with non-integer $N$, we write $(-1)^{N} = {\rm Re}[e^{i\pi N}]=\frac{e^{i\pi N}+e^{-i\pi N}}{2}$, the simplest form correct for integer $N$ (the
harmonic approximation violates this) and consistent with the
requirement that ${\cal G}_\sigma(x,\tau)$ be real and even in $x$. As a result, the Greens function takes the form ${\cal G}_\sigma(x,\tau)={\cal G}_{\sigma,+}(x,\tau)+{\cal G}_{\sigma,-}(x,\tau)$, with ${\cal G}_{\sigma,+}(x,\tau)=[{\cal G}_{\sigma,-}(x,\tau)]^*$.
Making the definitions,
$\Phi(x,\tau) \equiv \phi(x,\tau)-\phi(0,0)$ and
$\Theta(x,\tau) \equiv \theta(x,\tau)-\theta(0,0)$, we use
the Gaussian action resulting from \eqref{eq:H_c} to move the averages to the exponent,
\begin{eqnarray}
{\cal G}_{\sigma,+}(x,\tau) &\sim & e^{-2k_F|x| \frac{\ln(2S+1)}{\pi}}e^{i2k_Fx}\nonumber \\& &\times
\langle e^{i\left(1+i\frac{\ln(2S+1)}{\pi}\right)\Theta(x,\tau)} e^{i\Phi(x,\tau)}\rangle \nonumber \\
&=&e^{-2k_F|x| \frac{\ln(2S+1)}{\pi}}e^{i2k_Fx} e^{-\frac{1}{2}\left( 1+i\frac{\ln(2S+1)}{\pi}\right)^2 \langle \Theta^2\rangle}\nonumber \\
& &\times e^{-\frac{1}{2}\langle \Phi^2\rangle } e^{-\left(1+i\frac{\ln(2S+1)}{\pi}\right)\langle \Phi \Theta \rangle }\;.
\label{eq:G_bos}
\end{eqnarray}
Standard computations from Eq.~(\ref{eq:H_c}) using the relation \eqref{eq:N_with_theta} to relate $\theta=\sqrt{2}\theta_c$ and $\phi=\phi_c/\sqrt{2}$ give $\langle
\Theta^2\rangle=K_c \ln(x^2+v_c^2\tau^2)$, $\langle \Phi^2\rangle =
\frac{1}{4K_c}\ln(x^2+v_c^2\tau^2)$ and $\langle \Phi \Theta \rangle =
\frac{1}{2} \ln\left(\frac{v_c\tau -ix}{v_c\tau+ix}\right)$.
Substituting these values into Eq.~(\ref{eq:G_bos}), we find 
\bea\label{eq:G_asym_x}
{\cal G}_\sigma(x,\tau)=\frac{C'e^{-2k_F |x| \frac{\ln(2S+1)}{\pi}}}{(x^2+v_c^2\tau^2)^{\Delta_{K_c}}}\hspace{2cm} \nonumber \\
\times \left(\frac{e^{i(2k_F x -\varphi_{K_c}^+)}}{v_c \tau-ix}+\frac{e^{-i(2k_F
x -\varphi_{K_c}^-)}}{v_c \tau+ix}\right),
\eea
where $C'$ is an undetermined constant.\footnote{\textcite{cheianov03} determined it for the special case of infinite strength zero range interactions.} The anomalous
exponent $\Delta_{K_c}$ determining the power law decay is given by
\begin{equation}\label{eq:D_K_c}
\Delta_{K_c}=\frac{1}{8K_c}+\frac{K_c}{2}\left(1-\left(\frac{\ln(2S+1)}{\pi}\right)^2\right) -\frac{1}{2}\;,
\end{equation}
and the additional phase factors coming from the $-\frac{\ln(2S+1)}{\pi}(\langle \Theta^2\rangle +
\langle \Phi \Theta \rangle)= -\frac{\ln(2S+1)}{\pi}\left(K_c\ln
  \left(x^2+v_c^2\tau^2\right) + \frac{1}{2}\ln\left(\frac{v_c\tau-ix}{v_c\tau+ix}\right)\right)$ piece in the exponent are
\bea
\varphi_{K_c}^\pm(x,\tau)=\frac{\ln(2S+1)}{\pi}\Biggl (K_c \ln(x^2+v_c^2\tau^2) \nonumber \\ 
\pm \frac{1}{2}
\ln \left(\frac{v_c\tau-ix}{v_c\tau+ix}\right) \Biggr)\;.
\label{eq:phases}
\eea
For the special case of spin $S=1/2$ fermions (such as electrons) the result \eqref{eq:G_asym_x} reduces to \eqref{eq:G_SILL} from the introduction.  In the case of infinite strength, zero range interactions [when $K_c=1/2$ \cite{Schulz:prl90}] Eqs.\eqref{eq:G_asym_x}-\eqref{eq:phases} reduce to the tour-de-force Bethe ansatz results of \textcite{cheianov03,Cheianov04}.

There are several features of \eqref{eq:G_asym_x} worth emphasizing.  First, note that the exponential decay (coming from the non-fluctuating part of $(2S+1)^{-|N|}$) puts the correlations of the spin-incoherent Luttinger liquid out of the Luttinger liquid universality class because the correlation functions do not contain exclusively power law decays.\footnote{I emphasize again that the limit $T \to 0$ has been taken, so this is an exponential decay at zero temperature.}  Second, note that no parameter of the spin Hamiltonian (aside from the actual value of the spin itself) appears in the correlation function.  This is an explicit example of the ``super universal'' spin physics in which all parameters (including the symmetry) of the spin Hamiltonian have dropped out. It is interesting that the actual value of the spin, $S$, sets the scale of decay in both the exponential factor and the power law piece (via $\Delta_{K_c}$).  Aside from these differences, the spin-incoherent Luttinger liquid Greens function looks similar to that of the Luttinger liquid \eqref{eq:G_LL} in so far as it also posses the right/left moving structure.  However, as we will see shortly, the tunneling density of states (which is derived from the single particle Greens function) of the spin-incoherent Luttinger liquid is remarkably different from that of the Luttinger liquid.  In order to investigate this in detail we need a more accurate form for  ${\cal G}_\sigma(x,\tau)$ at small $x$.  We turn to this now.

\subsubsection{${\cal G}_\sigma(x,\tau)$ for small $x$}

Having discussed the spatial asymptotics of ${\cal G}_\sigma(x,\tau)$, for
$|x|\rightarrow\infty$, we now turn our attention to the small $x$ limit which will allow us to compute the low energy tunneling density of states at a
point (when $x=0$). Unlike the situation with $x \rightarrow \infty$, when
computing the Green's function at small $x$ one should be careful to take into 
account the discreteness of the the number of world lines that may 
``bend'' in between (0,0) and $(x,\tau)$ \cite{Fiete:prb05}:
\begin{eqnarray}\label{eq:G_smallx}
{\cal G}_\sigma(x,\tau)\sim \sum_{m=-\infty}^\infty (2S+1)^{-|m|}(-1)^m \hspace{2cm} \nonumber \\
\times \langle \delta(N(x,\tau)-m)e^{i[\phi(x,\tau)-\phi(0,0)]}\rangle \nonumber \\
= \sum_{m=-\infty}^\infty  (2S+1)^{-|m|}(-1)^m \hspace{2cm}  \nonumber\\ 
\times \int \frac{d\lambda}{2\pi} e^{-i\lambda m}
\langle e^{i[\lambda \Theta(x,\tau)/\pi + \Phi(x,\tau)]}\rangle \nonumber \\
\approx \sqrt{\frac{\pi}{2 \langle\Theta^2\rangle}} \sum_{m=-\infty}^\infty  (2S+1)^{-|m|}(-1)^m \hspace{.5cm}  \nonumber\\ 
\times e^{\frac{-\pi^2(\bar n x-m)^2}{2 \langle\Theta^2\rangle}} e^{-\langle \Phi^2\rangle/2}\nonumber \\
= \frac{a}{\sqrt{2\pi}\bar u(\tau)} \left(\frac{a}{v_c\tau}\right)^{\frac{1}{4K_c}} f\left(x, \bar u(\tau)\right) \;,\hspace{1cm}
\end{eqnarray}
where
\begin{equation}
\label{fdef}
f(z,\bar u) \equiv \sum_{m=-\infty}^\infty 2^{-|m|}(-1)^m  e^{-\frac{(x-ma)^2}{2 \bar u^2}}\;,
\end{equation}
and $\bar u=\frac{a}{\pi}\sqrt{\langle \Theta^2\rangle}$.
When $x=0$ we find \cite{Fiete:prl04}
\begin{eqnarray}
{\cal G}_\sigma(0,\tau)\sim \frac{1}{\sqrt{\tau^{\frac{1}{2K_c}}{\rm ln}(v_c\tau)}} \sum_{m=-\infty}^\infty (2S+1)^{-|m|}(-1)^m \nonumber \\
\times e^{-\frac{\pi^2 m^2}{4 K_c \ln (v_c\tau)}}\nonumber \\
\sim \frac{1}{\sqrt{\tau^{\frac{1}{2K_c}}{\rm ln}(v_c\tau)}},\hspace{3cm}
\label{eq:G_0}
\end{eqnarray}
where the final result \eqref{eq:G_0} is obtained by noting that the sum over $k$ depends only weakly on $\tau$ and ranges between $S/(S+1)$ and 1. This again recovers the Bethe ansatz results of \textcite{cheianov03,Cheianov04} for the case of infinite strength zero range interactions where $K_c=1/2$ \cite{Schulz:prl90}.  We will see momentarily that \eqref{eq:G_0} directly gives the tunneling density of states.

Before we leave the discussion of the single-particle Greens function, it is important to emphasize why I have divided the physics of the correlation functions into those derived from ``particle non-conserving'' and ``particle conserving'' operators.  The difference between these two types of operators (and their correlation functions) appears in the evaluation of the trace over the spin degrees of freedom {\em in the spin-incoherent regime} for which we used the ``non-crossing'' approximation to obtain a dominant contribution from trajectories that ``wrap around'' the imaginary time torus like those shown in Fig.~\ref{fig:worldlines}(a).  This class of trajectories is only possible for operators that create an ``end point'' of a world line, and these are precisely the operators that change particle number. These trajectories are responsible for the $(2S+1)^{-|N(x,\tau)|}$ factors, and all the associated spin-incoherent effects such as the exponential decay with distance of the single-particle Greens function \eqref{eq:G_asym_x} and the logarithmic time dependence in \eqref{eq:G_0}.  For a particle number conserving operator it is not possible to simultaneously statisfy the non-crossing approximation and have ``wrap around'' trajectories.  In fact, in Sec.~\ref{sec:cons} we will see that for quantities derived from particle conserving operators there is a precise mapping in the spin-incoherent regime between a spin-incoherent Luttinger liquid and a {\em spinless} Luttinger liquid. We have already seen hints of this in the spinless topology of the world line configurations after tracing over the spin degrees of freedom and in the operator relations between $\theta$ and $\theta_c$ in \eqref{eq:N_with_theta}.

A related issue to when it is possible to have ``wrap around'' trajectories is what happens to the form of the Greens function in the spin-incoherent regime when a particle is added near the boundary of the system, such as at the end of a semi-infinite wire.  At the end of a wire, the density fluctuations are effectively frozen out so that $N(0,\tau)\equiv 0$.  From the first line of \eqref{eq:G_smallx} this implies that only the $m=0$ term in the sum can contribute,
\be
\label{eq:G_end}
{\cal G}_{\sigma, {\rm end}}(0,\tau)\sim \langle e^{i[\phi(0,\tau)-\phi(0,0)]}\rangle_{\rm end} \sim \frac{1}{\tau^{\frac{1}{2K_c}}},
\ee
which compared to \eqref{eq:G_0} has a ``doubled'' exponent and the logarithmic correction characteristic of the spin-incoherence is absent.  Up to a factor of two in the exponent, \eqref{eq:G_end} is identical to the result that one would obtain for a {\em spinless} Luttinger liquid.  It is our second hint that the spin-incoherent Luttinger liquid may be related to a {\em spinless} Luttinger liquid in some ways.  Results for the Greens function near to, but not precisely at the end of a boundary are discussed in \textcite{Kindermann_crossover:prb06,Kakashvili:cm06,Fiete:prb05}.

\subsubsection{Tunneling density of states}
\label{sec:tdos}

One of the most remarkable differences between a spin-incoherent Luttinger liquid and a Luttinger liquid is the frequency (energy) dependence of the tunneling density of states.  It is well-known \cite{Giamarchi,Voit:rpp95,Glazman:prb92,Gogolin} that the tunneling density of states in a Luttinger liquid is suppressed as a power law \eqref{eq:tdos_LL} at low energies due to orthogonality catastrophe-type effects resulting from the interactions: when a new particle is added to the system the others must rearrange themselves to accomodate the additional particle.  The resulting final state wavefunction is orthogonal to the original one in the limit of large particle numbers.  While the physics of the orthogonality catastrophe is still operational in the spin-incoherent Luttinger liquid, the large number of highly excitated spin states a low energy turns out to dramatically affect the tunneling density of states.

The tunneling density of states is obtained \cite{Fiete:prl04,cheianov03,Cheianov04} by Fourier transforming \eqref{eq:G_0},
\bea
A(\omega)\sim \omega^{(1/4K_c)-1}/\sqrt{|\log(\omega)|},\;\;\;\; \hbar \omega \gtrsim k_B T.
\label{eq:tdos_SI}
\eea
Note that while the exponent is positive for any value of $K_c$ in \eqref{eq:tdos_LL}, for the spin-incoherent case the exponent is only positive if $K_c <1/4$. The value of $K_c$ for the infinite $U$ limit of the Hubbard model is 1/2 \cite{Schulz:prl90} and to obtain a smaller $K_c$ one needs a combination of both very strong and longer ranged interactions.  As a result, many systems will likely have $K_c>1/4$ and the tunneling density of states will exhibit a {\em divergence} at low energies for $\hbar \omega \gtrsim k_B T$.  This is qualitatively distinct from the Luttinger liquid and should be a relatively simple feature to observe in experiment.  The divergence in \eqref{eq:tdos_SI} has a straight forward interpretation: Orthogonality catastrophe-type physics operational in the charge sector competes with a highly degenerate spin sector at very low energies.  For $K_c >1/4$ the huge availability of spin states at low energies ``beats'' the orthogonality catastrophe and leads to an apparent divergence for $\hbar \omega \gtrsim k_B T$, while for $K_c <1/4$, the interactions in the charge sector are sufficiently strong to create an orthogonality catastrophe that overwhelms the degenerate spin states and a power law suppression of the tunneling density of states is recovered. 

\begin{figure}[h]
\includegraphics[width=.75\linewidth,clip=]{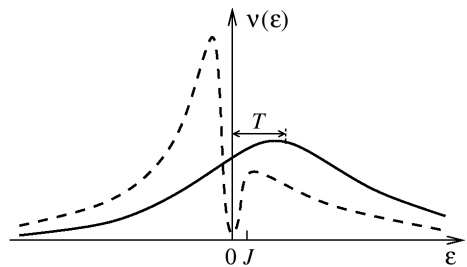}
\caption{\label{fig:Matveev_tdos} Schematic of the tunneling density of states, $\nu(\varepsilon)\equiv A(\omega)$ from \textcite{Matveev:prl06}. The solid (dashed) line is the high (low) temperature regime $k_B T \gg J$ ($k_BT \ll J$).  In the high temperature regime when $\varepsilon \gg k_B T$, $\nu(\varepsilon)=2\nu(-\varepsilon)$.  When the temperature is lowered below the spin energy $J$, the tunneling density of states at $\varepsilon<0$ grows by roughly a factor of 3, while for $\varepsilon>0$ it decreases dramatically.  For $k_BT \ll\varepsilon$ and  $\varepsilon \ll J$ the standard Luttinger liquid behavior of the power law suppression is observed at small $\varepsilon$.}
\end{figure}

The result \eqref{eq:tdos_SI} has been refined and extended to negative frequencies by \textcite{Matveev:prl06}.  In the spin-incoherent regime, the tunneling density of states possesses an asymmetry $A(\omega)=2A(-\omega)$.  The behavior is shown schematically in Fig.~\ref{fig:Matveev_tdos}.

\subsubsection{Finite magnetic field}

The spin-incoherent Luttinger liquid can be studied in a straight forward way in the presence of an external magnetic field. I will assume for this part of the discussion that there are no significant orbital effects from the external field.  This is certainly true if the system is strictly one dimensional, but real systems such as quantum wires are quasi-one dimensional and orbital effects may be important for sufficiently large fields. Neglecting any orbital coupling, the external field $B$ only couples to the spin of the particles (assumed $S=1/2$ here for simplicity), 
\be
H_Z=-g_e\mu_B B \sum_l S^z_l,
\ee
where $g_e$ is the gyromagnetic ratio of the electron and $\mu_B$ is the Bohr magneton.  

In the absence of a magnetic field, the Greens functions satisfy ${\cal G}_\uparrow(x,\tau)={\cal G}_\downarrow(x,\tau)$. But this is no longer the case in the presence of a magnetic field.  Nevertheless, all of the arguments used in the beginning of Sec.~\ref{sec:Greens_gen} to evaluate the trace over the spin degrees of freedom using the ``non-crossing'' approximation still hold as they are based only on  $E_{\rm spin} \ll k_BT$.  The only change is that now the probability of having spin projection $\sigma$ along the z-axis depends on $\sigma$ and the value of the external field.  In particular, the factor $(2S+1)^{-|m|}$ in \eqref{eq:G_gen} gets modified to\footnote{We have actually dropped an unimportant overall factor of $(2S+1)^{-1}$ in \eqref{eq:G_gen}, but now that we are discussing finite magnetic fields we must include this additional factor in the form of its finite field generalization $p_\sigma$.  The probability of having $N$ spins aligned is $(2S+1)^{-|N|}$, but the probability of having $N$ spins alinged in a {\em particular direction} is $(2S+1)^{-1}\times(2S+1)^{-|N|}$, giving the extra mutliplicative factor.} $p_\sigma^{|m|+1}$ with $p_\uparrow=(1+{\rm exp}\{-E_Z/k_BT\})^{-1}$ and $p_\downarrow=1-p_\uparrow$, where $E_Z=2g_e\mu_B  B$ is the Zeeman energy of an electron in a magnetic field referenced to the minimum energy configuration with the spin parallel to the field.

All of the calculations presented so far carry through as before only with the changes above.  \textcite{Kindermann_crossover:prb06} have studied these effects on the single-particle Greens function and the tunneling density of states.\footnote{External magnetic field effects on the Fermi edge singularity \cite{Fiete:prl06} and the momentum structure in momentum resolved tunneling \cite{Fiete:prb05} have also been studied.}  The main result is that a new time scale
\be
\tau_B =\frac{a}{v_c}e^{E_Z/(K_c k_BT)},
\ee
is introduced which sets a cut off for spin-incoherent effects.  For times much longer than this spin-incoherent Luttinger liquid effects are observed, while for times much shorter than this, the system behaves like a spinless Luttinger liquid.  In the limit of very large external fields, $E_Z\to \infty$, and therefore $\tau_B \to \infty$, there is no time (frequency) range over which spin-incoherent Luttinger liquid physics may be observed. This result, of course, squares with intuition: A {\em fully} polarized spin-incoherent Luttinger liquid should be identical to a spinless Luttinger liquid.
This is indeed the case.

\subsection{Fermi-edge singularity}

\begin{figure}[h]
\includegraphics[width=.8\linewidth,clip=]{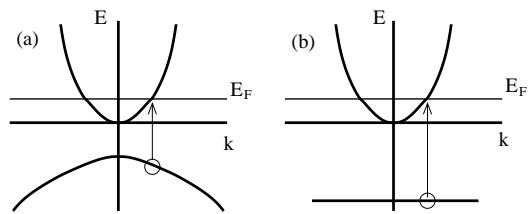}
\caption{\label{fig:absorption} Schematic of threshold photo-excitation for (a)
finite hole mass and (b) infinite hole mass.  In (a) an electron is excited from the valence band to the conduction band leaving behind a hole in valence band.
 In (b) an electron is excited from a deep (infinite mass) core level.  We are assuming the ``conduction band" is occupied by a spin-incoherent Luttinger liquid.}
\end{figure}

In the Fermi-edge singularity, an incoming photon excites a deep ``core'' level electron up to the Fermi energy as shown schematically in Fig.~\ref{fig:absorption}.  There is a minimum energy required to do this\footnote{In two and three dimensions, this statement is only strictly true if the core level is infinitely massive.} called the threshold energy and the energy dependence of the photon absorption just above the threshold energy is what constitutes the Fermi-edge singularity, as it is often singular. The detailed form of the singularity can reveal a great deal about the interactions in the system, the mass of the core hole created, and the absence or presence of spin-incoherent degrees of freedom \cite{Fiete:prl06}.   

The photon absorption rate is computed by Fermi's Golden rule and is given by
\be
\label{eq:threshold}
I(\omega) \propto \sum_\sigma {\rm Re}\int_0^\infty dt e^{i\omega t} \langle \psi_\sigma(t)h_\sigma(t)h_\sigma^\dagger(0)\psi_\sigma^\dagger(0)\rangle,
\ee
where as before the operator $\psi_\sigma^\dagger$ ($\psi_\sigma$) creates (annihilates) an electron and $h_\sigma^\dagger$ ($h_\sigma$) creates (annihilates) a hole with spin $\sigma$.  The expression \eqref{eq:threshold} shows that the photon absorption is related to a correlation function derived from particle non-conserving operators, $h_\sigma^\dagger\psi_\sigma^\dagger$, and its Hermitian conjugate. We therefore expect spin-incoherent effects to be manifest.

The result of analysis \cite{Fiete:prl06} is that the Fermi-edge singularity falls into two classes of behavior depending on whether the hole created by the photon is localized (infinitely massive) or delocalized (finite mass). The important distinction between the two cases is that the former breaks the translational symmetry of the system while the latter does not.  The most significant consequence of the breaking of translational symmetry is that backscattering from an infinitely massive impurity is relevant (in a system with repuslive interactions) and cuts the electron system into two semi-infinite parts \cite{Kane:prl92,Furusaki:prb93}, while backscattering from a finite mass impurity is irrelevant \cite{Neto:prb96}.  These two limits have important implications for the boundary conditions the $\theta$ and $\phi$ operators appearing in $\psi_\sigma$ satisfy.\footnote{Recal from our discussion earlier that the Greens function for tunneling into a point \eqref{eq:G_0} in an infinite systems is different from the Greens function for tunneling into the end \eqref{eq:G_end} of a semi-infinite system because of the different boundary conditions on $\theta$.  As before, the spinless $\theta$ and $\phi$ fields are related to the spinful ones via the relation described in \eqref{eq:N_with_theta} and in the sentence just below it.}

In the finite hole case shown in Fig.~\ref{fig:absorption}(a) one can transform to a frame co-moving with the excited hole \cite{Neto:prb96,Tsukamoto:prb98,Tsukamoto:epj98}.  In this frame the Hamiltonian takes the form $H=H_{\rm elec}+H_{\rm elec-hole}+H_{\rm hole}$, where
\be
H_{\rm elec-hole}=\frac{U_s^f}{\pi} h^\dagger h \partial_x\theta(0) \pm  \frac{U_a^f}{\pi}h^\dagger h \partial_x\phi(0)\,,
\ee
$H_{\rm elec}=H_c+H_s$,   $H_{\rm hole}=\sum_\sigma E_{h,\sigma} h_\sigma^\dagger h_\sigma$ and $h^\dagger h=\sum_\sigma h_\sigma^\dagger h_\sigma$.  Here $U_s^f$ is the symmetric part of the forward scattering from the hole and $U_a^f$ is the antisymmetric part of the forward scattering \cite{Tsukamoto:prb98}.
(In our convention $\partial_x\theta$ represents the density fluctuations and $\partial_x\phi$ the particle current.)  The antisymmetric part appears since in the frame of the
hole, it sees a net current of particles scattering from it. The ``+'' sign is for a right-moving hole and the ``-'' sign is for a left-moving hole. The parameter $U_a^f$ depends on the momentum and mass of the hole, and when it is at rest,  $U_a^f\equiv 0$ \cite{Tsukamoto:prb98}. Since the backscattering from a finite mass impurity is not relevant it has
no affect on the Fermi-edge physics and therefore has not been included in $H_{\rm elec-hole}$.  The Hamiltonian $H$ can be diagonalized with the unitary transformation $U={\rm exp}\{-i[\delta_a\theta(0)+\delta_s\phi(0)]h^\dagger h\}$ where $\delta_a\equiv \mp U_a^f/(v_cK_c\pi)$ and $\delta_s \equiv -2K_cU_s^f/(v_c\pi)$.
Applying this transformation we find $\bar H  \equiv U^\dagger H U =  H_{\rm elec}+\bar
H_{\rm hole}$, where the only change to $H_{\rm hole}$ is a shift in the hole energy $E_{h,\sigma} \to \tilde E_{h,\sigma}$. The correlation function appearing in \eqref{eq:threshold} can now be computed in imaginary time, $C_\sigma(\tau)=\frac{1}{Z}{\rm Tr}\left[e^{-\beta H}  \psi_\sigma(\tau)h_\sigma(\tau)h_\sigma^\dagger(0)\psi_\sigma^\dagger(0)\right]
=\frac{1}{Z}{\rm Tr}\left[e^{-\beta \bar H}  \bar \psi_\sigma(\tau)\bar h_\sigma(\tau)\bar h_\sigma^\dagger(0)\bar \psi_\sigma^\dagger(0)\right],$ where in the second line $U U^\dagger=1$ has been inserted and the cyclic property of the trace has been exploited.  Direct evaluation gives $\bar \psi_\sigma = \psi_\sigma$ (up to unimportant multiplicative factors) and $\bar h_\sigma=h_\sigma e^{-i[\delta_a\theta+\delta_s\phi]} $, so that
\bea
%\label{eq:C}
C_{\sigma}(\tau)=\frac{e^{-\tilde E_{h,\sigma}\tau}}{Z_{\rm elec}}{\rm Tr}\biggl[e^{-\beta H_{\rm elec}}   \psi_\sigma(\tau) e^{-i[\delta_a\theta(\tau)+\delta_s\phi(\tau)]}  \nonumber \\
\times e^{i[\delta_a\theta(0)+\delta_s\phi(0)]} \psi_\sigma^\dagger(0)\biggr],
\eea
which is now of exactly the same form as the Greens function \eqref{eq:G_def}.  Carrying through the same manipulations outlined in Sec.~\ref{sec:Greens_gen},
we find
\be
\label{eq:I_SI}
I(\omega) \propto (\omega-\omega_{\rm th})^{\frac{1}{4K_c}(1-\delta_s)^2-1}/\sqrt{|\ln(\omega-\omega_{\rm th})|}
\Theta(\omega-\omega_{\rm th}),
\ee
where $\Theta(\omega-\omega_{\rm th})$ is the step function and $\omega_{\rm th}=\tilde E_{h,\sigma}/\hbar$ is the threshold frequency.  A few of the most important features of \eqref{eq:I_SI} are worth emphasizing.  First, in contrast to the spin coherent (spin polarized) Luttinger liquid \cite{Tsukamoto:prb98,Tsukamoto:epj98} the threshold exponent {\em does not depend on the mass of the core hole}.  Second, there are ``universal" (independent of interactions) logarithmic corrections to the power-law threshold behavior.  These logarithmic corrections are of the same nature as those that arose in the tunneling density of states in the spin-incoherent Luttinger liquid, \eqref{eq:tdos_SI}.  Third, there is indeed a minimum (threshold) frequency for absorption.

The case of infinite hole mass shown in Fig.~\ref{fig:absorption}(b) is qualitatively different because now the hole is a relevant perturbation.  At frequencies just above the threshold frequency $\omega_{\rm th}$, the hole acts just as if it were an ``end'' in the system \cite{Kane:prl92,Furusaki:prb93}.  As a result, the effective low-energy boundary condition is $\theta(\tau)=0$, just as we found for the Greens function at the end of a semi-infinite system, \eqref{eq:G_end}. The Fermi-edge singularity therefore maps onto an equivalent spinless problem.  It has been shown that for a spinless Luttinger liquid the infinitely massive core hole leads to a universal back scattering contribution to the exponent of 1/8 \cite{Furusaki:prb97,Komnik:prb97,Prokofev:prb94,Gogolin:prl93} and this leads to the infinite mass threshold result  $I(\omega) \propto (\omega-\omega_{\rm th})^{\frac{1}{2K_c}(1-\delta_s)^2+1/8-1}\Theta(\omega-\omega_{\rm th})$. As with the Greens function at the end of a semi-infinite system, there is no logarithmic correction to the frequency dependence. For symmetry reasons mentioned before there is non-analytical behavior in the exponent when the infinite mass limit is taken for the hole \cite{Castella:prb96}.

\subsection{Spin-incoherent effects in transport}

Transport experiments in mesoscopics are by now fairly routine and provide fertile grounds for observing spin-incoherent effects.  In this subsection we summarize some of the results from the theory of the spin-incoherent Luttinger liquid that can be readily probed in transport experiments.

\begin{figure}[h]
\includegraphics[width=.95\linewidth,clip=]{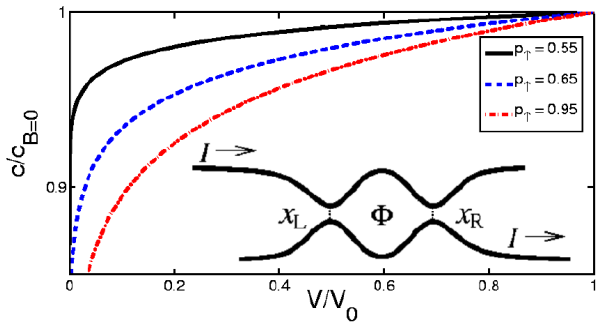}
\caption{\label{fig:Kindermann_interference} (color online)  Interference schematic in the inset.  The interference contrast \eqref{eq:C_def} obeys a power law with a temperature and magnetic field dependent exponent $2K_c(1-\ln^2p_\uparrow/\pi^2)$, Eq.\eqref{eq:C}.  Plotted is $c=C(V)/C(V_0)$, the interference contrast normalized with respect to an arbitrary reference voltage $V_0$.  It is compared to its value at zero magnetic field $c_{B=0}$ for various polarizations $p_\uparrow$ with $K_c=1/2$. From \textcite{Kindermann:prl06}.}
\end{figure}

Spin-incoherent effects are expected to have dramatic consequences on interference experiments such as those shown schematically in the inset of Fig.~\ref{fig:Kindermann_interference}. This situation has been studied in detail by \textcite{Kindermann:prl06} and \textcite{LeHur:prb06}. In this type of interference experiment two quantum wires are in close proximity and tunneling dominates at two spatial points, $x_L$ and $x_R$.  A flux $\Phi$ penetrates the region between these wires bounded by $x_L$ and $x_R$ as indicated in the figure.  A current $I$ is injected in the top wire to the left of $x_L$ and the resulting current is measured in the lower wire to the right of $x_R$.  For the current in the lower wire to be finite, electrons must tunnel from the upper wire either at $x_L$ or $x_R$.  In general, there will be a non-zero amplitude for each and depending on the flux $\Phi$ there will be either constructive or destructive interference (provided the two paths are coherent) leading to an oscillating current as a function of $\Phi$.
 
A covenient way to quantify the coherence of the system is with the interference contrast $C$,
\be
\label{eq:C_def}
C=\frac{\sqrt{\langle(I(\varphi)-\langle I\rangle_\varphi)^2\rangle}}{\langle I\rangle_\varphi},\;\;\;\langle \dots \rangle_\varphi = \int_0^{2\pi}\frac{d\varphi}{2\pi}.
\ee
\textcite{Kindermann:prl06} showed that in the spin-incoherent regime the contrast behaves as
\be
\label{eq:C}
C\sim p_\uparrow(eV)^{2K_c(1-\ln^2p_\uparrow/\pi^2)},
\ee
where $p_\uparrow=(1+e^{-E_Z/k_BT})^{-1}$ is the probability of having a spin up electron.  The interference contrast exhibits an anomalous scaling with with voltage, magnetic field, and temperature.  Numerical results are presented in Fig.~\ref{fig:Kindermann_interference}.

So far, we have delt only with topics that involve infinite or semi-infinite sytems.  In any real transport situation a quasi-one dimensional system will ultimately be connected to Fermi liquid leads and this will affect many aspects of transport.  For example, the dc conductance of an infinite single mode spinless Luttinger liquid with interaction parameter $g$ is $g \frac{e^2}{h}$ \cite{Kane:prl92}, while if Fermi liquid leads are attached the dc conductance becomes $\frac{e^2}{h}$ independent of $g$ \cite{Safi:prb95,Maslov:prb95}. It is therefore important to know how the finite length of the wire and the Fermi liquid leads will affect the observation of spin-incoherent Luttinger liquid physics.

\begin{figure}[h]
\includegraphics[width=.7\linewidth,clip=]{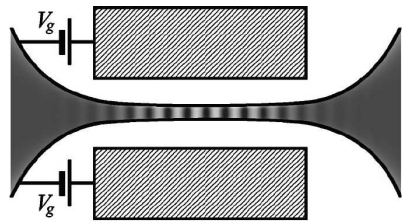}
\caption{\label{fig:Matveev_wire}  Schematic of a quantum wire attached to Fermi liquid leads.  The electron density is assumed to be small near the center of the wire so that $r_s$ is large and strong Wigner solid correlations are present. In the leads to the left and right of the wire the electrons are assumed to be non-interacting. From \textcite{Matveev:prb04}.}
\end{figure}

Poineering work in this direction was done by \textcite{Matveev:prl04,Matveev:prb04} in the context of the $0.7(2e^2/h)$ conductance feature \cite{Thomas:prl96,Cronenwett:prl02,DiCarlo:prl06} in quantum point contacts.  Matveev modeled the quantum point contact as a finite length quantum wire of very low particle density adiabatically connected to Fermi liquid leads as shown in Fig.~\ref{fig:Matveev_wire}.  The conductance was computed by using a Wigner solid model like that given in Eq.\eqref{eq:H_s} and Eq.\eqref{eq:H_c} where $J_l$ was assumed to be spatially dependent.  For electrons near the center of the wire $J_l\ll E_F$, while it grew to of order $E_F$ in the leads.  Remarkably, when the temperature of the system is such that the wire is in the spin-incoherent regime, $J \ll k_B T \ll E_F$, the dc conductance of the wire is reduced to  half the non-interacting value, $0.5(2e^2/h)$, which is close to the $0.7(2e^2/h)$ feature.  One physical interpretation of this result is that when an electron in the leads with energy $\sim k_BT$ enters the constricted region of the wire it starts to decompose into separate spin and charge components.  Since the bandwith of the spin modes scales as $J_l \ll k_B T$ there are no propagating spin modes in the wire at the energy of the electron and these states are all reflected while the charge modes are allowed to pass through.  This argument explains the reduction of the conductance.  The factor of 2 reduction (relative to the non-interacting value of $2e^2/h$) is more subtle and appears to depend on the symmetry of the spin Hamiltonian in the wire \cite{Matveev:prl04,Matveev:prb04}. Recently this issue has also been addressed numerically \cite{Syljuasen:cm06}. Interestingly, in the regime $J \ll k_B T \ll E_F$ it appears the shot noise can also be dramatically reduced in certain situations \cite{Kindermann_noise:prb06}. 

\begin{figure}[h]
\includegraphics[width=.8\linewidth,clip=]{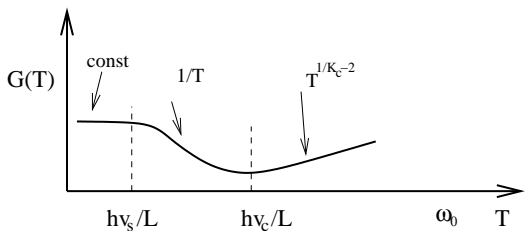}
\caption{\label{fig:cond_temp_regimes} Schematic of the temperature dependence of a quantum wire of length $L$ attached to Fermi liquid leads with a strong impurity in the center.  For a range of temperatures $\hbar v_s/L \ll k_B T \ll \hbar v_s/L$ the charge modes have propagated into the leads, but the spin modes have not leading to a ``universal'' temperature dependence of the conductance in this energy window.}
\end{figure}

If a strong impurity is added to the center of a wire like that shown in  Fig.~\ref{fig:Matveev_wire}, interesting non-monotonic behavior of the conductance $G(T)$ may result \cite{Fiete_2:prb05}.  For a strong impurity, the conductance can be calculated using \eqref{eq:G_end} with Fermi's Golden rule for tunneling across the impurity.  The result is schematically shown in Fig.~\ref{fig:cond_temp_regimes}.  For $k_B T \gg \hbar v_c/L, \hbar v_s/L$ the quantum wire behaves just like an infinite system because the excitations do not ``see'' the leads.  When $ \hbar v_s/L \ll k_B T \ll \hbar v_c/L$ the charge modes ``see'' the leads but the spin modes do not.  Naively taking the non-interacting value of $K_c=1$ leads to an inverse temperature dependence down to temperatures $k_B T \lesssim \hbar v_s/L$ where the system ``looks'' like a strong impurity in a Fermi liquid and the value of the conductance saturates.

%\subsection{n-body operators}
%SC pair correlations. universal decay here.  Describe any of the results from new project?

\section{Particle conserving operators}
\label{sec:cons}

In some sense, the greatest qualitative differences between a spin-incoherent Luttinger liquid and a Luttinger liquid are seen in the correlation functions derived from the particle non-conserving operators that we have been discussing up to this point.  This is because of the particular class of trajectories possible [see Fig.~\ref{fig:worldlines}(a)] for particle non-conserving operators in the ``non-crossing'' approximation.  Upon evaluating the trace over the spin degrees of freedom, factors such as $(2S+1)^{-|N|}$ (and its finite magnetic field generalization $p_\sigma$) appear, leading to exponential decays in space \eqref{eq:G_asym_x} and logarithmic corrections in time \eqref{eq:G_0}.  As I mentioned before, such trajectories are not possible for particle conserving operators.  Nevertheless, there are important ways in which spin-incoherent effects manifest themselves for this class of operators, namely in the temperature dependence of the correlation functions.

For this section I will focus on only one particle conserving operator--the density.  All of the physics discussed here will be related to the weak $2k_F$ density modulations in the large $r_s$, strongly interacting limit.  For $k_B T \ll E_{\rm spin}$ these $2k_F$ density modulations will be present due to the weak spin-charge coupling, while for $k_B T \gg E_{\rm spin}$ they will be thermally washed out.  Therefore, quantities depending on the $2k_F$ density modulations, such as the Coulomb drag between parallel quantum wires and certain types of noise, will exhibit a sharp temperature dependence around $k_B T \approx E_{\rm spin}$.  Aside from the physics summarized in Fig.~\ref{fig:cond_temp_regimes} in which a spin energy ($\hbar v_s/L$) explicitly appeared, up to this point all of the physics we have discussed has explicitly or implicity taken $E_{\rm spin}\to 0$.  In this section, it will be important to keep $E_{\rm spin}$ finite in order to compute the $2k_F$ density modulations.  However, before we discuss that in detail, one more result is in order for the $E_{\rm spin} \to 0$ limit.  We turn to this now.

\subsection{Mapping to spinless electrons in the spin-incoherent Luttinger liquid}

In this subsection we show explicitly that in the spin-incoherent Luttinger liquid
electrons become effectively
spinless (for quantities that do not directly probe spin and do not change particle number) and are
governed by a Hamiltonian of the form (\ref{eq:H_c}) with interaction
parameter $g=2K_c$ \cite{Fiete_2:prb05}. 

To show this microscopically, we work in the canonical
ensemble, i.e. with a fixed number of electrons. The dynamics
in the grand canonical ensemble can be obtained from this by summing
over the sectors with each electron number.  For fixed electron number,  
a convenient real-space basis set is given by
states specifying the position $x_l$ of each electron, and the spin
projection on the $\hat{z}$ axis, $\sigma_l$, in order, from left to right
across the system:
\begin{equation}
  \label{eq:fockbasis}
  |x_1\cdots x_N\rangle |\sigma_1\cdots \sigma_N\rangle =
  c_{\sigma_1}^\dagger(x_1)\cdots c_{\sigma_N}^\dagger(x_N)|0\rangle,
\end{equation}
where $|0\rangle$ is the vacuum state (no particles).

As we discussed in the context of the evaluation of the single-particle Greens function, the physics of the spin-incoherent regime is that, within the
thermal coherence time, $t_{coh} \sim \hbar/k_B T$, the probability of
a transition between states with different values of $\{ \sigma_n \}$ is
negligible.  Hence, the physics is well-approximated by neglecting
off-diagonal matrix elements in these states.  Moreover, in the same
approximation, for spin-independent interactions, the matrix
elements of $H$ are independent of the $\{ \sigma_n \}$, 
\begin{eqnarray}
  \label{eq:matelts}
&& \langle \sigma'_1\cdots \sigma'_N|\langle x'_1\cdots x'_N|H|x_1 \cdots x_N\rangle
|\sigma_1\cdots \sigma_N\rangle \nonumber \\
&& \approx \langle x'_1\cdots x'_N|H_{\rm spinless}|x_1 \cdots
x_N\rangle \delta_{\sigma'_1,\sigma_1}\cdots \delta_{\sigma'_N,\sigma_N}\nonumber
\end{eqnarray}
where $H_{\rm spinless }$ is an effective spinless Hamiltonian that governs the (independent) dynamics
within each spin sector which must have the same {\em form} as \eqref{eq:H_c}.  It only remains to determine the mapping between the coupling contants of $H_c$ and $H_{\rm spinless }$.  To do this we equate a physical quantity between the two representations, such as the density.  In fact, we have already done this in Eq.~\eqref{eq:N_with_theta} which shows $\theta=\sqrt{2}\theta_c$ and $\phi=\phi_c/\sqrt{2}$. Upon substitution into \eqref{eq:H_c}, this gives an identical collective mode velocity but a new coupling constant $g=2K_c$ \cite{Fiete_2:prb05}.  This equivalence is completely general and continues to hold in the presence of arbitrary potentials, weak links, etc, so long as $J\ll k_B T$ throughout the system, there are no explicit spin-dependent interactions in the
Hamiltonian, and electrons are not added or removed from the
system during the dynamics.  The single-particle Greens function for the infinite system \eqref{eq:G_asym_x} does not satisfy that last condition and that is why it does not map onto the Greens function of some spinless Luttinger liquid.

\subsection{Spin-charge coupling in Wigner solid model}

In the limit of large $r_s$, the dominant finite wavevector spatial correlations are at $4k_F$ corresponding to an average spacing $a$ between particles, as schematically shown in Fig.~\ref{fig:solid_schematic_rev}. Recall that for a non-interacting electron gas the dominant finite wavevector spatial correlations are at $2k_F$ where $k_F \equiv \pi/(2a)$.  The difference between the two is simply understood because for non-interacting electrons two particles (a spin up and a spin down) can occupy the same position in space, while for large $r_s$ the interactions are so strong that two particles tend not to be near each other in space regardless of their relative spin orientation.

Returning for the moment to $T=0$, one may ask how the Wigner solid ``crosses over'' to the non-interacting limit in 1-d as a function of $r_s$.\footnote{Because of the strong quantum fluctuations in one dimension, a true Wigner solid does not exist.  However, the longer range the interactions, the more slowly the density correlations decay. It turns out that for Coulomb interactions, the correlations decay {\em slower} than any power law \cite{Schulz:prl93}.}  Starting from the fluctuation Wigner solid model \eqref{eq:H} one can address this question by allowing coupling of the spin and charge degrees of freedom.  In particular, the $2k_F$ oscillations (characteristic of the weakly interacting case) can be seen to be brought in through the $l$-dependence of the coupling $J_l$ between spins in \eqref{eq:H_s}.  Assuming that the local fluctuations from the equilibrium positions $u_l$ are small compared to the mean particle spacing $a$, the exchange energy can be expanded as \cite{Fiete:prb06},
\begin{equation}
J_l=J +J_1(u_{l+1}-u_l) + {\cal O}((u_{l+1}-u_l) ^2)\;.
\end{equation}
In this case the full Hamiltonian takes the form
\begin{equation}
\label{eq:H_wire}
H=H_c + H_s(J_l=J) + H_{s-c}\;,
\end{equation}
where
\begin{equation}
\label{eq:H_m-e}
 H_{s-c}= J_1\sum_l (u_{l+1}-u_l){\vec S_{l+1}}\cdot{\vec S_l} \;.
\end{equation}
Eq.\eqref{eq:H_m-e} explicitly couples the spin modes to the elastic distortions of the lattice constituting
the charge modes.  The displacement $u_l$ of the $l^{th}$ electron in the harmonic chain \eqref{eq:H_chain} and \eqref{eq:H_m-e} can be expanded as 
\be
\label{eq:u}
u_l=u_0(la)+u_\pi(la)(-1)^l,
\ee
where $u_0$ refers the  $k\approx 0$ component of the displacement and
$u_\pi$ refers to the $k\approx \pi/a\equiv 2k_F$ displacement.  Both $u_0$ and $u_\pi$
are assumed to be slowly varying functions of position, and we expect
$u_\pi \ll u_0$ when the interactions are strong.  When \eqref{eq:u} is substituted into \eqref{eq:H_m-e} and the continuum limit is taken, it is the $u_\pi$ ($ 2k_F$) piece that will couple to the spin degrees of freedom.  Since the spin sector is expected to have antiferromagnetic correlations (which therefore oscillate at wavevector $2k_F$), this is the crucial coupling that will allow the $2k_F$ correlations of the spin sector to ``spill over'' into the charge sector.  Of course, there are only correlations in the spin sector provided $k_BT \ll E_{\rm spin}$.  In this way the presence or absence of $2k_F$ correlations in the charge density of a strongly interacting 1-d system can tell one about the relative size of $k_BT$ and  $E_{\rm spin}$.

\subsection{Description of the effective density} 

This idea can by quantified by starting with the density representation $\rho(x,t)=\sum_l \delta(x-al -u_l(t))$. The density can be expanded in the displacements $u_l$, and the higher energy $u_\pi$ pieces can be integrated out to yield an effective density valid when $k_BT \ll E_{\rm spin}$ \cite{Fiete:prb06}
\bea
 \label{eq:rho_eff_final}
\rho^{\rm eff}(x,\tau) = \rho_0 -\frac{\sqrt{2}}{\pi}\partial_x\theta_c(x,\tau)\hspace{3.5 cm}\nonumber \\ -
\frac{\rho_0}{16\pi}\left(\frac{J_1}{m\omega_0^2 a^2}\right)
\sin\left(2k_Fx+\sqrt{2}\theta_c(x,\tau)\right)\sin(\sqrt{2}\theta_s(x,\tau))
\nonumber \\
+ \rho_0 \cos\left(4k_Fx+\sqrt{8}\theta_c(x,\tau)\right),\hspace{.5 cm}
\eea
where $\rho_0=1/a$.

The density expression \eqref{eq:rho_eff_final} has effectively traded the high energy (because they are at the zone boundary) $u_\pi$ modes in favor of the spin variables $\theta_s$ described by \eqref{eq:H_s_low}.  
Note that the prefactor $\left(\frac{J_1}{m\omega_0^2 a^2}\right)\sim\frac{v_s}{v_c}\sim \frac{E_{\rm spin}}{E_{\rm charge}}$ of the $2k_F$ piece is non-universal and depends on the strength of the interactions.\footnote{The prefactor of the $2k_F$ contribution to \eqref{eq:rho_eff_final} may depend also on the doping as explicitly seen in the Hubbard model \cite{Giamarchi}.}  Since the spin-incoherent regime requires $E_{\rm spin} \ll E_{\rm charge}$, evidently the $2k_F$ oscillations are weak even at zero temperature.

\subsection{$2k_F$ density correlations}

The $2k_F$ part of the density correlations (which show the most important temperature dependence for spin-incoherent effects) can be computed in a straight forward way for $k_B T \ll E_{\rm spin}$ using \eqref{eq:H_c} and \eqref{eq:H_s_low} \cite{Fiete:prb06,Fiete:prb07,Iucci:prb07}.  The main result is that these correlations are {\em exponentially} suppressed when $k_B T \gtrsim E_{\rm spin}$. Therefore, in order to understand the effects of spin-incoherent physics one may compute the $T=0$ result within the low energy theory \eqref{eq:H_c} and \eqref{eq:H_s_low} to determine what will be {\em missing} when  $k_B T \gtrsim E_{\rm spin}$.\footnote{Interesting changes are also expected in the same temperature range for the momentum distribution \cite{Cheianov:cm04}.}

\subsection{Coulomb drag, noise, and dephasing}

In this subsection we discuss ways that the loss of $2k_F$ oscillations may be observed in experiment. Spin-incoherent effects are revealed through the temperature dependence of the $2k_F$ density correlations in: (i) Coulomb drag between quantum wires, (ii) the voltage noise on a gate close to a wire, and (iii) the dephasing time of a qubit near a wire. In many cases, unusual non-monotonic temperature dependence may result providing a clear signature of the spin-incoherent Luttinger liquid.

\begin{figure}[h]
\includegraphics[width=.5\linewidth,clip=]{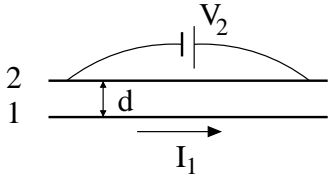}
\caption{\label{fig:drag} Coulomb drag schematic.  Two quantum wires are arranged parallel to one another and separated by a distance $d$.  A dc current $I_1$ flows in the ``active'' wire 1 and a voltage bias $V_2$ is measured in the ``passive'' wire 2 when $I_2=0$.}
\end{figure}

\begin{figure}[h]
\includegraphics[width=.8\linewidth,clip=]{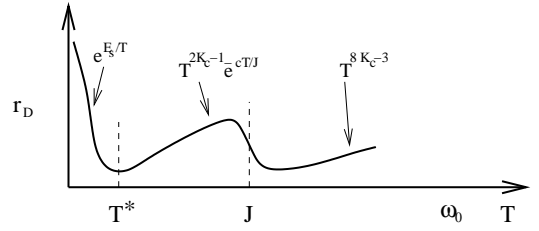}
\caption{\label{fig:temp_regimes} Schematic of the possible temperature dependence of the Coulomb drag in indentical wires of sufficiently low electron density that $J \ll E_{\rm charge}=\hbar \omega_0$.  The non-monotonic temperature dependence shown can be obtained for $K_c>1/2$ and $\tilde U(4k_F) \ll \tilde U(2k_F)$ which may be realized when $k_Fd>1$.  The
temperature $T^*$ is the ``locking" temperature of two identical wires, below which the drag exhibits activated behavior and $E_s$ is an energy gap associated with the ``locking". 
 When $J\ll E_{\rm charge}$ a sharp drop in the drag resistance should be observable for $K_B T \sim J$.}
\end{figure}

In the Coulomb drag experiment (see Fig.\ref{fig:drag}) a current $I_1$ is driven in an ``active'' wire while a voltage bias $V_2$ is measured in a ``passive'' wire.  The drag resistivity is defined as $r_D =-\lim_{I_1 \to 0} \frac{e^2}{h}\frac{dV_2}{dI_1}$ and can be expressed \cite{Pustilnik:prl03,Zheng:prb93} in terms of the imaginary part of the Fourier transformed retarded density-density correlation function $\chi^{R}(k,\omega)$  computed from  \eqref{eq:rho_eff_final} \cite{Fiete:prb06},
\be
\label{eq:drag_Pustilnik}
r_D=\int_0^\infty dk \int_0^\infty d\omega \frac{k^2  \tilde U^2_{12}(k)}{4 \pi^2 n_1 n_2 T} \frac{{\rm Im}\chi^R_1(k,\omega) {\rm Im}\chi^R_2(k,\omega)}{\sinh^2(\omega/2T)}\;,
\ee
where $n_i$ is the density of wire $i$ and $\tilde U_{12}(k)$ is the Fourier transform of the {\em interwire} interaction.  The structure of \eqref{eq:rho_eff_final} leads to $\chi(k,\omega)\approx \chi^{k\approx 0}(k,\omega)+\chi^{2k_F}(k,\omega)+\chi^{4k_F}(k,\omega)$, so that the drag resistivity results from a sum of three terms. With a harmonic charge theory \eqref{eq:H_c}, the $k\approx 0$ contribution generically vanishes \cite{Pustilnik:prl03,Fiete:prb06} leaving the drag to be dominated by the $2k_F$ and $4k_F$ pieces of $\chi(k,\omega)$.  Since  $\tilde U_{12}(2k_F) \gg \tilde U_{12}(4k_F)$ when $k_F d>1$, it is possible for the drag to be determined by the $2k_F$ contributions at low temperature.  However, when $k_B T \gg E_{\rm spin}$ these correlations will be absent leaving only a $4k_F$ contribution to the drag.  The resulting temperature dependence is shown in Fig.~\ref{fig:temp_regimes} \cite{Fiete:prb06}.

\begin{figure}[h]
\includegraphics[width=.80\linewidth,clip=]{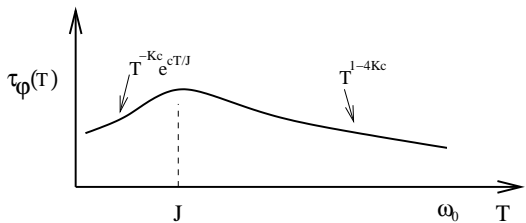}
\caption{\label{fig:temp_dephasing} Schematic temperature dependence of the dephasing time of a qubit.  The constant $c$
in the exponential is ${\cal O}(1)$. The figure assumes $1/4 <K_c <1$ and $K_s=1$.}
\end{figure}

If a metallic gate is placed in proximity to quantum wire in which a finite current $I$ is driven, there will be frequency-dependent voltage noise given by
$S_Q(\omega)=\frac{1}{2}\int dt e^{i\omega t} \langle \{V(t),V(0)\}\rangle$, where $V(t)$ is the voltage on the gate.  Relating the voltage fluctuations to the density fluctuations in the wire gives \cite{Fiete:prb07} 
\be
\label{eq:noise}
S_Q(\omega)=\int \frac{dq}{2\pi} |u(q)|^2 \frac{\chi(q,\omega)+\chi(q,-\omega)}{2},
\ee
where $S_Q(\omega)= S_Q^{q\approx 0}(\omega)+S_Q^{2k_F}(\omega)+S_Q^{4k_F}(\omega)$ from the structure of $\chi(q,\omega)$.  Here $u(q)$ is the Fourier transform of the interaction between the gate and the wire.  The resulting voltage noise at zero temperature is \cite{Fiete:prb07}
\bea
S_Q^{q\approx 0}(\omega)&\propto&|u(0)|^2 |\omega|,\nonumber\\
S_Q^{2k_F}(\omega)&\propto&\left(\frac{J_1}{m\omega_0^2 a^2}\right)^2\sum_\pm |u(2k_F)|^2 \left|\omega \pm \frac{\omega_I}{2}\right|^{K_s+K_c-1}\nonumber\\
S_Q^{4k_F}(\omega)&\propto& \sum_\pm |u(4k_F)|^2|\omega \pm \omega_I|^{4K_c-1},
\eea
where $\omega_I=2\pi I/e$. The frequency dependence of the voltage noise measured thus displays power law singularities at the frequencies $\omega_I/2$ and
$\omega_I$ that are observable at low  temperatures.
The singularity in $S_Q^{2k_F}$ at $\omega\approx \omega_I/2$, however, becomes  exponentially small as $k_B T \gtrsim J$ thus indicating an entry to the spin-incoherent regime.  The dephasing time $\tau_\varphi$ of a qubit is determined by the zero frequency part of the noise and is shown as a function of temperature in Fig.~\ref{fig:temp_dephasing} \cite{Fiete:prb07}.

\section{Experimental evidence for spin-incoherence}

As the previous sections suggest, the theory of the spin-incoherent Luttinger liquid has developed rapidly in recent years.  There are now many falsifiable predictions testable in experiment that are quite distinct from those of the low temperature Luttinger liquid. On the experimental side much remains to be explored.  So far, little directed effort has been made to investigate spin-incoherent effects, but this situation is already changing rapidly. 

At present there are strong indications in momentum resolved tunneling experiments on cleaved-edge overgrowth quantum wires \cite{Auslaender:sci05,Steinberg:prb06} that the spin-incoherent regime $E_{\rm spin} \ll k_BT \ll E_{\rm charge}$ is indeed obtained at large $r_s$ (small electron densities).  Unfortunately, the analysis is somewhat involved \cite{Fiete:prb05,Steinberg:prb06}. There is certainly urgent need for experiments that probe the spin-incoherent Luttinger liquid in more direct ways.  My goal in this section is to present the strongest (in my view) experimental evidence that this state is indeed reached in high quality ballistic semiconductor quantum wires at low particle density.\footnote{Besides the momentum-resolved tunneling experiments discussed here, there is perhaps also the $0.7 (2e^2/h)$ conductance features in quantum point contacts that could be explained by spin-incoherent effects in a finite length quantum wire attached to Fermi liquid leads \cite{Matveev:prl04,Matveev:prb04}.}

\subsection{Momentum resolved tunneling at low particle density}

\begin{figure}[h]
\includegraphics[width=.9\linewidth,clip=]{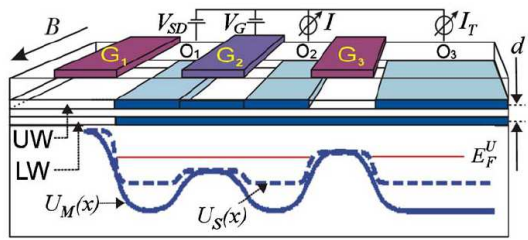}
\caption{\label{fig:Amir_schematic} (color online)  A schematic of the measurement set up with cleave plane front and perpendicular to the magnetic field $B$.  The top gates ($G_1,G_2,G_3$) are 2 $\mu$m wide.  The upper wire at the edge of the 2DEG is 20 nm thick, the lower wire is 30 nm thick, and the barrier between them is 6 nm insulating AlGaAs. Here $U_S(x)/U_M(x)$ is the gate-induced potential for the single-mode/multi-mode wires, $E^U_F$ is the Fermi energy of the upper wire, $O_1$ is an ohmic contact that serves as a source, and the ohmic contacts $O_{2,3}$ serve as drains.  The electron density in the quantum wires is modulated by a gate voltage $V_G$. The tunneling current is $I_T$ and the two-terminal current is $I$. From \textcite{Steinberg:prb06}.}
\end{figure}

In many tunneling experiments, such as those done with a scanning tunneling microscope (STM), the tunneling essentially occurs at a single point--at the STM tip. For an STM, it can be shown that the tunneling conductance is related to the local density of states of the surface~\cite{Tersoff:prb85,Fiete:rmp03}.  For a fixed tip position, there is only energy resolution in the density of states.  There is no momentum resolution  because a local state must be built up of many different momentum states.  Momentum resolved tunneling is achieved when the tunneling occurs at many {\em equivalent} points, such as when there is translational symmetry in the system.  This is precisely the case for tunneling between two {\em parallel} quantum wires.  The translational symmetry between parallel wires guarantees that momentum will be conserved when particles tunnel from one wire to the other.  By applying a magnetic field {\em perpendicular to the plane of the wires} as shown in Fig.~\ref{fig:Amir_schematic} a {\em tunable momentum boost proportional to the field strength and independent of the energy} can be given to the tunneling particles.  The energy, as with the STM, can be adjusted with a source-drain voltage bias.  Thus, momentum resolved tunneling gives an additional ``knob'' (the momentum) to adjust when studying tunneling between two translationally invariant systems.  As is well known \cite{Mahan}, to lowest order in the tunneling, the current is given by the spectral functions $A(k,\omega)$ of the two systems (the upper and lower wire in the present case).  Therefore, with independent control of both momentum and energy, the spectral functions of the double-wire system can be extracted.  Having   $A(k,\omega)$ in hand is tantamount to knowledge of the full dynamical properties and excitations of the system.  This is just the information needed to observe dynamical effects such as spin-charge separation,  and has in fact been obverved in this type of set up~\cite{Auslaender:sci02,Auslaender:sci05}.

A simple argument shows how the momentum boost occurs.  Define a coordinate system so that the $x$ direction is parallel to the quantum wires (labeled by UW/LW for upper wire/lower wire) shown in Fig.~\ref{fig:Amir_schematic}.  Let the $y$-direction be perpendicular to the $x$-direction and {\em parallel} to the plane containing the wires, and let $z$ be the direction of the magnetic field which is {\em perpendicular} to the plane of the wires. It is straightforward to see that ${\vec B}={\vec \nabla}\times {\vec A}$, where ${\vec A}=(0,Bx,0)$.  Consider an electron moving in the lower wire with a wavefunction proportional to $e^{ikx}$ that tunnels to the upper wire.  After tunneling in the presence of the magnetic field it will pick up an Ahronov-Bohm type phase equal to $\frac{e}{\hbar c}\int {\vec A}\cdot d{\vec l}=\frac{e}{\hbar c}\int_0^d Bx dy=q_B x$, where $q_B=\frac{eBd}{\hbar c}$ is the momentum boost and $d$ is the center-to-center distance between the quantum wires as shown in Fig.~\ref{fig:Amir_schematic}.  Thus, upon tunneling the state $e^{ikx}\to e^{i(k+q_B)x}$, which looks like a momentum boost.  As claimed earlier, the momentum boost is proportional to the magnetic field and independent of the energy.  For an interacting system in which the eigenstates are not plane waves, the eigenstates may still be expressed as a linear combination of plane wave states.  Since the momentum boost $q_B$ is independent of $k$, each plane wave state will have the same boost so that the overall state will have the same boost.  Therefore, regardless of the state of the electrons in the upper and lower wires there will be a momentum boost of $q_B$ upon tunneling. 

\begin{figure}[h]
\includegraphics[width=.9\linewidth,clip=]{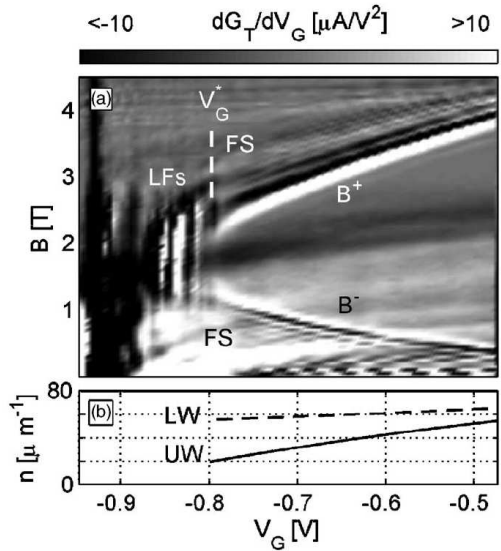}
\caption{\label{fig:Amir_localization} Experimental data indicating a localization transition to a strongly correlated regime at low particle density. (a) Plot of $dG_T/dV_G$ vs $V_G$ and $B$ for a single mode quantum wire.  The applied bias $V_{SD}=100 \mu$V is selected to avoid the zero-bias anomaly, but is still small enough to allow tunneling predominantly between the Fermi points of both wires.  The upper and lower curves are momentum conserving tunneling features $B^\pm(V_G)$ and each curve is accompanied by finite size features marked by ``FS''.  At low densities localized features (LFs) appear instead of these curves. The localization transition occurs at $V^*_G$.  (b) indicates the upper wire (UW) and lower wire (LW) densities extracted from the data. From \textcite{Steinberg:prb06}.}
\end{figure}

The type of experimental data that indicates the spin-incoherent physics is shown in Fig.~\ref{fig:Amir_localization}. To acquire this data the source drain voltage is set to a fixed value, $V_{SD}=100 \mu$V, which is small enough to keep the tunneling in the linear response regime.  The tunneling conductance $G_T$ is then measured as a function of the perpendicular magnetic field $B$, and the voltage $V_G$ which modulates the voltage on gate $G_2$ shown in Fig.~\ref{fig:Amir_schematic}.  When the gate $G_2$ is not grounded, the upper wire is effectively divided into three regions: (1) The region between the right edge of $G_1$ and the left edge of $G_2$, (2) The region directly underneath $G_2$, and (3) The region between the right edge of $G_2$ and the left edge of $G_3$. Because all of these regions of the upper wire are biased by $V_{SD}$ relative to the lower wire, electrons tunnel from all three regions.  The main goal of the experiments reported by \textcite{Steinberg:prb06} was to study the low density regime of the quantum wires by applying a large negative bias to $G_2$.  This creates an effective potential $U_{M/S}(x)$ shown schematically in Fig.~\ref{fig:Amir_schematic} that leads to a low density region in the center of the upper wire.  In order to isolate the contribution to tunneling of this central, low density region of the upper wire an effort was made to ``subtract off'' the contributions from the two higher density ``end'' regions.  Under the assumption that the end regions are unaffected by small changes in $V_G$, the tunneling contribution from the electrons in the central portion can be isolated by examining the change in $G_T$ with $V_G$, that is $dG_T/dV_G$.  This type of data is shown in Fig.~\ref{fig:Amir_localization}.  A brighter signal indicates larger tunneling.

\begin{figure}[h]
\includegraphics[width=.9\linewidth,clip=]{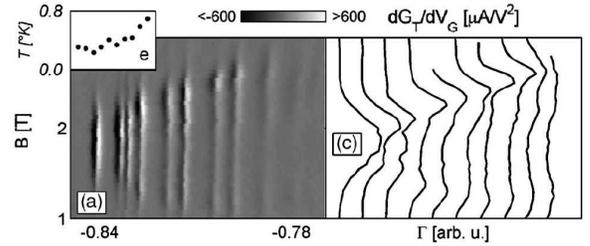}
\caption{\label{fig:Amir_LF} Detail of the localization features in the strongly correlated regime at low particle density. (a) A high resolution measurement of $dG_T/dV_G$ of localization features for a single-mode wire, $V_{SD}=50\mu$V, $dV_G=300\mu$V. (e) $T(N)$ of panel (a) where $N$ is the number of added electron for each LF. (c) $\Gamma(B)\propto |\Psi(k)|^2$. $\Gamma$ and $T$ are defined in the text.From \textcite{Steinberg:prb06}. }
\end{figure}

There are several important features of the data presented in Fig.~\ref{fig:Amir_localization}.  First, for $V_G>V_G^*$ there are two values of magnetic field, $B^+$ and $B^-$, at any particular $V_G$, where the largest tunneling occurs.  This is the expected result for electrons which are delocalized and freely propagating.  For example, it can be shown~\cite{Steinberg:prb06,Auslaender:sci05} that the physics of the $B^-$ signal is electrons tunneling from right (left) moving states in one wire to right (left) moving states in the other.  In other words they are  ``forward tunneling events''.  On the other hand, the $B^+$ signal corresponds to electrons tunneling from right (left) moving states in one wire to left (right) moving states in the other, i.e. ``backward tuneling events'', where I have borrowed the language from forward and backward scattering.  In order for this ``forward'' and ``backward'' interpretation to make sense, the electronic states must be reasonably close to states where momentum is a good quantum number, as it is for free electrons.  The sharpness of the $B^-$ and $B^+$ for $V_G > V_G^*$ indicates this is in fact the case.\footnote{This is not to say that the system is not a Luttinger liquid at these values of $V_G$.  In fact, a careful analysis of the data for these $V_G$ values show clear signatures of spin-charge separation and other Luttinger liquid effects~\cite{Auslaender:sci05}. The spin and charge velocities can even be determined as a function of electron density in the wires. }

The second remarkable feature is the behavior of the data for $V_G<V_G^*$. Over this range of gate voltages the two distinct $B^-$ and $B^+$ features go away and are replaced by very broad (in magnetic field) features.  These broad features, called localized features (LFs) in Fig.~\ref{fig:Amir_localization}, are also discrete in gate voltage indicating the onset of Coulomb blockade behavior below $V_G^*$.  As we will show in the Sec.~\ref{sec:theory_exp}, the localized features have an interpretation in terms of strong Wigner solid-like correlations and excited spin degrees of freedom. 

The localized features are shown in more detail in Fig.~\ref{fig:Amir_LF}(a).  The intensity of the localized features can be quantified by the ``weight'', $\Gamma$, under a given localized feature for a specified $B$ value~\cite{Steinberg:prb06}.\footnote{As explained in \textcite{Steinberg:prb06}, $\Gamma$ and $T$ are fit from the Coulomb blockade peaks to the formula $\partial^2f/\partial V_G^2=(e^2/h)\Gamma \pi/4k_BT^2\tanh(x)\cosh^2(x)$ where $x=[(\epsilon-\mu)/k_BT]$.}  The weight $\Gamma$ is shown in  Fig.~\ref{fig:Amir_LF}(c) and the temperature in Fig.~\ref{fig:Amir_LF}(e).  The crucial feature for us is the the double-lobed structure of $\Gamma$ as a function of $B$.

Before I turn to the theory of these experiments, it is important to emphasize the most important features of the data in Fig.~\ref{fig:Amir_LF}. First, for gate voltages more negative than those shown no more peaks appear.  This indicates that for voltages just slightly larger (to the right) of the first peak there is one electron in the central region of the wire. Likewise, for voltages just larger than the value of the second peak, there are two electrons in the central portion of the wire, and so on. Second, the integrated (over $B$) weight under the peaks tends to decrease as the particle number increases.  As we discuss in the next subsection, this effect is related to the ``orthogonality catastrophe''. Third, for the second peak and higher, $\Gamma$ has a double lobed structure.  The data is slightly asymmetric, and this probably related to an asymmetric effect of $G_2$ on the electron density in the central region of the upper wire. 

It is the {\em details}\footnote{This reliance on details is the primary reason why there is an urgent need for more direct experimental probes of the spin incoherent regime. The theory is now sufficiently well developed to predict many ``smoking gun'' signatures in a variety of different experiments.  It is my hope that this article will help motivate such experiments to explore this intriguing regime of strongly interacting one dimensional systems.} of the double lobed $\Gamma$ that reveal the likelihood that $E_{\rm spin} \ll k_B T \ll E_{\rm charge}$ is reached at the low particle (density) limit in the quantum wires. The most important features of the double lobed structure are that the separation of the peaks grows with particle number, the widths of the peaks are comparable to the separation between them, and the center ``dip'' between the peaks is shallower than the tails off to the sides.  That the spacing and widths of the peaks track the particle number smoothly indicates that a collective, many-body effect is responsible for the shape.  This behavior {\em would not occur if the electrons were all falling into distinct local minima}.  It turns out that the separation between the peaks is related to the $k_F$ of the upper wire, and the widths of the peaks as well as the shallow ``dip'' in between them all have a very natural interpretation interms of a fluctuating Wigner solid model with highly excited spin degrees of freedom.  It is the purpose of the next subsection to support this claim with direct theory that {\em quantitatively} suggests a fluctuating Wigner solid model is reasonable and  $E_{\rm spin} \ll k_B T$ is indeed reached for the relevant experimental parameters.

\subsection{Theoretical support for the spin-incoherent Luttinger liquid in experiment}
\label{sec:theory_exp}

\begin{figure}[h]
\includegraphics[width=.8\linewidth,clip=]{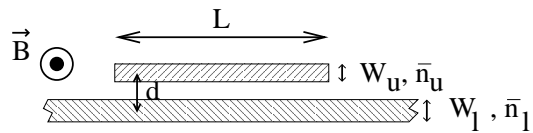}
\caption{\label{fig:geometry} Schematic geometry of electron tunneling
between two parallel quantum wires, as in Fig.~\ref{fig:Amir_schematic}. Electrons are assumed to tunnel between an
infinitely long lower wire and a short upper wire of length $L$.  The wires are
separated by a center-to-center distance $d$.  The upper (lower) wire has
a width $W_u$ ($W_l$) and an average electron density $\bar n_u$ ($\bar n_l$).}
\end{figure}

The main objective of this section is to provide theoretical support for the claim that a fluctuating Wigner solid model with thermally excited spins can explain the {\em details} of the doubled lobed structure of $\Gamma(B)$ in Fig.~\ref{fig:Amir_LF}(c).  In order to avoid complicating our discussion of the data any more than necessary we consider a slightly simpler situation than the one shown in Fig.~\ref{fig:Amir_schematic}.  Our model isolates the central, low density region of the upper wire just under $G_2$ and neglects the two ``ends'' between the gates $G_1$ and $G_3$. This geometry is shown in Fig.~\ref{fig:geometry}.

\subsubsection{Exact diagonalization for few electrons}
 
\begin{figure}[h]
\includegraphics[width=.8\linewidth,clip=]{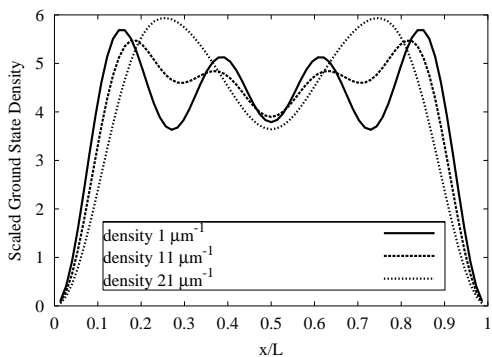}
\caption{\label{fig:density_spin} Ground state density of a
four electron system
    with spin.  Scaled densities are measured in units of $1/L$.
        In the label of
        the individual plot, ``density'' means the \emph{average
        physical} density $N/L$ of the system, in units of $\mu\rm{m}^{-1}$.
}
\end{figure}

As a first step towards quantitatively understanding the experiments of \textcite{Steinberg:prb06}, one would like to accurately study the few electron limit of the upper wire.  \textcite{Fiete:prb05} carried out exact diagonalizations of four electrons in the upper wire assuming the geometry in Fig.~\ref{fig:geometry}, but using realistic interactions that took into account the widths of the quantum wires, $W_u/W_l$, the separation between them, $d$, and the dielectric constant of the material. The confining potential of the upper wire was assumed to be infinite at $x=0$ and $x=L$.  The results for the ground state density are shown in  Fig.~\ref{fig:density_spin}.    Note the physical density dependence of the oscillations with position  $x$ along with wire.  For $n\approx 10 e/\mu$ there is a transition from predominantly $2k_F$ oscillations at higher densities to predominantly $4k_F$ oscillations at lower densities.  In fact, for $n\approx 1 e/\mu$ one can essentially ``see'' where the four electrons are sitting along with wire.  What these exact numerics tell us is that for densities lower than $n\approx 10 e/\mu$ the ground state electron density starts to look very much like that of a Wigner solid.  Returning for a moment to the experiments of~\textcite{Steinberg:prb06}, the best estimate of the density at which the ``localization transition'' occurs is $n\approx 10-20 e/\mu$.  Taken together, these result suggest that the transition may be associated with a tendency towards strong Wigner solid correlations in the depleted region of the upper wire.  I again stress that the gate voltage dependence of the peaks in $\Gamma(B)$ shown in Fig.~\ref{fig:Amir_LF}(c) are {\em inconsistent with electrons falling into distinct local minima}.

\begin{figure}[h]
\includegraphics[width=.9\linewidth,clip=]{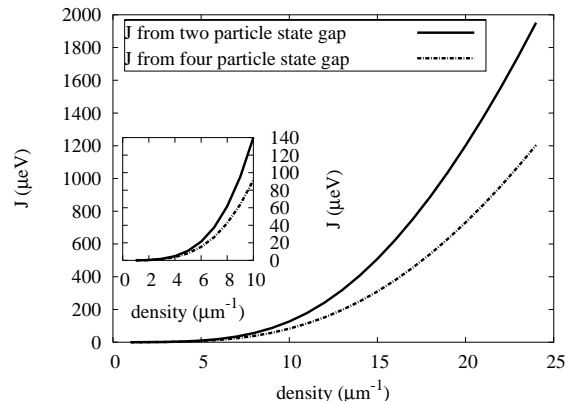}
\caption{\label{fig:gap_comp} Estimate of the spin exchange energy $J$ from a two and four electron system as a function of density. Typical experimental temperatures ($T_{\rm exp}$) are in the range 250 mK-2 K $\approx 25  \mu $eV-200 $\mu$eV.}
\end{figure}

Within the exact diagonalizations for the four electron system, the spin states can also be studied.  As expected from the Lieb-Mattis theorem \cite{lieb62}, the ground state is antiferromagnetically ordered.  In the low density limit, the numerical results showed \cite{Fiete:prb05} that indeed the spin system can be approximated by the Heisenberg spin chain, \eqref{eq:H_s}.  The spin excitation energies can also be studied by flipping one spin away from the antiferromagnetic configuration and asking how the energy cost depends on the density of the electrons.  From this, the nearest neighbor exchange $J$ can be computed.  The results are shown in Fig.~\ref{fig:gap_comp}.  Returning again to experiment, for densities $n \lesssim 10-20 e/\mu$ the upper wire is in the ``localized regime'' and characterized by strong Wigner solid correlations.  It is useful to compare the computed exchange energy $J$ with the temperature of the experiment, $T_{\rm exp}$.  One finds that for $n \lesssim 10 e/\mu, J<k_B T_{\rm exp}$, suggesting that at the lowest densities the spin incoherent regime should be accessible. 

So far, the numerics have provided evidence that near the localization transition, Wigner solid-like correlations are developing and the magnetic exchange energy may be of order the experimental temperature or smaller, opening the possibility of a description in terms of the model introduced in Sec.~\ref{sec:Wigner}.  While these estimates suggest the possibility of the spin-incoherent regime at the lowest densities, the strongest evidence comes from the {\em detailed} line shape of $\Gamma(B)$.  We therefore require a theory for this quantity.

The experimental results in Fig.~\ref{fig:Amir_localization}(b) indicate that the gate $G_2$ preferentially depletes the upper wire, leaving the density of the lower wire relatively unchanged and also with a much larger density relative to the upper wire.  Since at zero gate voltage both the upper wire and the lower wire are in the high density regime where interactions are less important, we can make an approximation that treats the lower wire as non-interacting throughout the full range of $V_G$.  In this approximation~\cite{Fiete:prb05}, the tunneling conductance on the Coulomb blockade peaks\footnote{In Coulomb blockade theory tunneling occurs when there is a degeneracy of an $N-1$ and $N$ particle state.} is $G_T=G_T^++G_T^-$ where $G_T^\pm \propto {\cal B}(k^\pm)$ with
\be
\label{eq:B_k}
{\cal B}(k(B))=\sum_{\alpha\gamma\sigma}|\langle \Psi_\alpha^N|c^\dagger_{k(B)\sigma}|\Psi_\gamma^{N-1}\rangle |^2.
\ee
Here $\Psi_\alpha^N$ is an $N$-particle eigenstate of the upper wire with quantum numbers $\alpha$ (perhaps the total spin and $z$-component of the spin), $c^\dagger_{k\sigma}$ creates a state with wavevector $k$ and $z$-component of the spin $\sigma$, and  $k^\pm(B)=\pm k_F^l +eBd/\hbar c$, where $k_F^l$ is the Fermi wavevector of the lower wire. The result \eqref{eq:B_k} tells us that the tunneling conductance on the localized feature is  proportional to the overlap of an $N$-particle eigenstate with a state which is an $(N-1)$-particle eigenstate plus a plane wave state.  If we define
\be
M(k(B))= \langle \Psi_\alpha^N|c^\dagger_{k(B)\sigma}|\Psi_\gamma^{N-1}\rangle,
\ee
then $M(k)$ can be expressed as the Fourier transform of a quasi-wavefunction,
\be
M(k(B))=\int dx e^{ik(B)x}\Psi^{N*}_{\rm eff}(x)/\sqrt{L},
\ee
where $\Psi^{N*}_{\rm eff}(x)=\langle \Psi_\alpha^N|\psi_{\sigma}(x)|\Psi_\gamma^{N-1}\rangle$.  Therefore, the magnetic field dependence of the conductance on the Coulomb blockadge peaks reveals the Fourier transform of an effective wavefunction \cite{Steinberg:prb06}.

\begin{figure}[h]
\includegraphics[width=.75\linewidth,clip=]{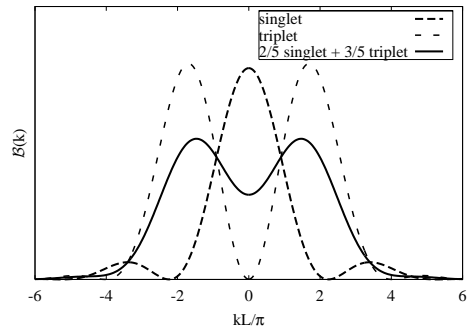}
\caption{\label{fig:FT12} Momentum-dependence of tunneling conductance between one-electron and
two-electron states for a wire of length $L$=0.4$\mu$m.  Dashed curve shows
the line-shape $|M_s(k)|^2$ for tunneling into the singlet ground
state of the two particle system.  The short-dashed curve shows $|M_t(k)|^2$,
for tunneling into the triplet ground state, applicable at $T=0$
when $E_Z >J$.  Solid curve is a weighted average, applicable if $k_BT$ is
large compared to both $E_Z $ and $J$ but small compared to the energy of the lowest charge excitation.  The singlet + triplet line shape should be compared with the $\Gamma(B)$ line shape in Fig.~\ref{fig:Amir_LF}(c).}
\end{figure}

The Fourier transform picture gives a satisfying interpretation of the momentum (magnetic field) structure and the orthogonality catastrophe.  If the electrons were non-interacting, there would a peak at $\pm k_F^u$, with a width $\sim 1/L$ due to the finite length  of the system and the $B$-integrated weight would be independent of $N$.  When interactions are present, adding a new electron shifts the states of all the electrons previously in the system leading to an orthogonality catastrophe.  This effect can explain the diminishing of the $B$-integrated weight of the localized features with increasing particle number $N$ in Fig.~\ref{fig:Amir_LF}(a\&c).  The most crucial point is that the double lobed structure with a shallow ``dip'' can naturally be explained in terms of a strongly interacting state with  highly thermally excited spin states.  An illustrative example is the case of $N=2$.  In this case, the ground state is a singlet.  A singlet ground state will have a symmetric orbital part of the wavefunction.  This leads to a maximum in $M(k)$ at $k=0$, as shown in Fig.~\ref{fig:FT12}.  If the state the electron tunneled into was instead a triplet state, the orbital state is antisymmetric and this leads to a zero in $M(k)$ at $k=0$, as shown in Fig.~\ref{fig:FT12}.  On the other hand, if the temperature is large, both singlet and triplet states are energetically allowed and this leads to the double lobed structure with a shallow dip in the middle.  The singlet + triplet line shape in Fig.~\ref{fig:FT12} should be compared with the $\Gamma(B)$ line shape in Fig.~\ref{fig:Amir_LF}(c).  As we will see in the next subsection, this feature is generic to a fluctuating Wigner solid model with highly thermally excited spins, that is, a spin-incoherent Luttinger liquid.

\subsubsection{Comparison with spin-incoherent Luttinger liquid theory}

In this subsection, we show the finite temperature momentum structure of ${\cal B}(k)$ in Fig.~\ref{fig:FT12} is generic to the spin-incoherent Luttinger liquid whose effective Hamiltonian is \eqref{eq:H}.  To do this we compute ${\cal B}(k)$ using the spin-incoherent Luttinger liquid theory.  This requires computing $A(k,\omega)$ of the upper wire which, as we already discussed earlier, is determined from the Greens function ${\cal G}(x,\tau)$.  Because we are interested in large $\tau$ (small energies) the Fourier transform will be dominated by contributions from $|x| <v_c \tau$. The dominant, short-distance, $x\sim a$, correlations that determine the momentum composition of the Green's function should thus be correctly described by Eq.~(\ref{eq:G_smallx}).  The Fourier transform is \cite{Fiete:prb05},
\begin{equation}
\label{GktA}
{\cal G}_\sigma(k,\tau) \sim a \left(\frac{a}{v_c\tau}\right)^{\frac{1}{4K_c}} A\left(k, \bar u (\tau) \right)\;,
\end{equation}
where
\begin{equation}
\label{Adef}
A(k, \bar u ) \equiv e^{-\frac{k^2\bar u^2}{2}} \left [\frac{3}{5+4 \cos(ka)}\right ]\;,
\end{equation}
and $\bar u(\tau) =\frac{a}{\pi}\sqrt{\langle \Theta(\tau)^2\rangle}$ which saturates at $\bar
u(\tau=L/v_c)=\frac{a}{\pi}\sqrt{2 K_c \ln(L/a)}$ for long times in a finite
system.

\begin{figure}[h]
\includegraphics[width=.7\linewidth,clip=]{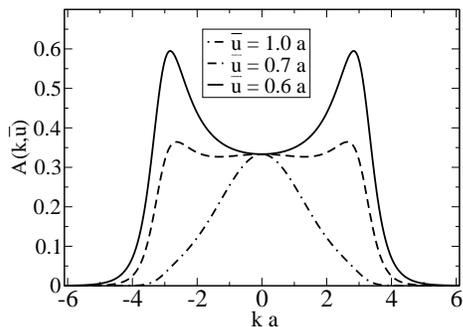}
\caption{\label{fig:smallx_FT_u} Spectral function $A(k, \bar u)$,
which determines the momentum dependence of tunneling in the spin-incoherent regime.  The quantity $\bar u$ is the
root-mean-square electron displacement, due to quantum fluctuations,
from the sites of a classical Wigner crystal, and $a$ is the lattice
spacing. When $\bar u \gtrsim a$, the momentum distribution is single
lobed and peaked about zero momentum.  In the opposite limit, when
$\bar u \ll a$, the momentum distribution exhibits a doubled lobed
structure with peaks near $k=\pm 2 k_F$.}
\end{figure}

Equations (\ref{GktA}) and (\ref{Adef}) are the central results  for
the spin-incoherent Luttinger liquid and they have several features worth emphasizing.  The first
is the momentum structure: There is an exponential envelope centered
about zero momentum, $e^{-\frac{k^2\bar u^2}{2}}$, whose width is
given by the parameter $\bar u$ measuring the fluctuations of an
electron's position.  Larger fluctuations imply a more sharply peaked
envelope in momentum space.  This envelope multiplies another momentum
dependent function, which is sensitive to the mean spacing of the
electrons and has maxima at $k=\pm \pi/a= \pm 2k_F$.  Results for
different $\bar u$ are shown in Fig.~\ref{fig:smallx_FT_u}.  These results should be compared with Fig.~\ref{fig:Amir_LF}(c) and the finite temperature results (singlet + triplet) of Fig.~\ref{fig:FT12}.  The double peaked structure with the soft ``dip'' is robust provided the fluctuations $\bar u$ are smaller than the interparticle spacing $a$, i.e., the system is a fluctuating Wigner solid with highly excited spins.

\section{Outlook and open issues}

In this colloquium, I have attempted to introduce the reader to the concept of the spin-incoherent Luttinger liquid and to explain how this state of strongly interacting 1-d systems fits in with the more familiar Luttinger liquid state. While there are some similarities, there are important differences which make it an intriguing field of study.  I have identified two different classes of correlation functions:  1) Those derived from particle non-conserving operators and 2) Those derived from particle conserving operators. In the first class, exponential decays in space and logarithmic dependence in time (and frequency) are ubiquitious features in the spin-incoherent regime.  In the second class, the correlation functions can be mapped onto a spinless Luttinger liquid deep in the spin-incoherent regime.  However, these correlation functions do show dramatic temperature dependence when $k_B T \approx E_{\rm spin}$ and we have discussed how this appears in Coulomb drag experiments between quantum wires and voltage fluctuations on a metallic gate near a quantum wire.  

A number of open theoretical issues remain. There are many details of the crossover from the Luttinger liquid regime to the spin-incoherent Luttinger liquid regime that are not well understood because they are difficult to address analytically.  Numerical work is thus highly desirable and may provide useful results for directly comparing to experiments, especially those which happen to be in the regime $k_B T \approx E_{\rm spin}$ where spin-incoherent effects are not fully manifest. Another issue that has not been addressed is how spin-orbit effects modify the spin-charge coupling discussed here and the energy scale for the observation of spin-incoherent effects. The coupling of external magnetic fields to orbital degrees of freedom should also be investigated.  There are also related systems, such as the edges of quantum Hall states that may exhibit ``incoherent'' effects, perhaps in a neutral sector of the edge modes.  1-d cold atomic gases may also provide a realization of some of the physics discussed here.

On the experimental side, I have tried to present some of the best indications to date of the spin-incoherent Luttinger liquid in high-quality quantum wires. Theoretical estimates of the relevant energy scales and direct calculation of observable quantities provide compelling evidence that the spin-incoherent Luttinger liquid has been observed in momentum resolved tunneling. Unfortunately, the analysis relies on a fairly detailed study of the data.  The theory has now been sufficiently developed that there are many more direct ways to probe the spin-incoherent Luttinger liquid.  Such experiments are what is most urgently needed in the field.  I sincerely hope that experimentalists will take up the challange.  It will undoubtedly lead to many insights and results not yet anticipated.

\acknowledgments
I am grateful and deeply indebted to all my collaborators on this subject: Ophir Auslaender, Leon Balents, Bert Halperin, Markus Kindermann, Karyn Le Hur, Jiang Qian, Hadar Steinberg, Yaroslav Tserkovnyak, Amir Yacoby, and to the countless others with whom I have had insightful discussions.  Among them are Matthew Fisher, Akira Furusaki, Thierry Giamarchi, Leonid Glazman, and Kostya Matveev.  This work was supported by NSF grants PHY05-51164, DMR04-57440, the Packard Foundation, and the Lee A. DuBridge Foundation.  I am also grateful for the hospitality of the Aspen Center for Physics where part of this work was done.

%\bibliography{fiete_nanowires.bib}

\end{document}